\documentclass[aps,prd,email,showpacs,showkeys,preprintnumbers,amsmath,amssymb,nofootinbib,superscriptaddress,twocolumn]{revtex4}
\usepackage{slashed}
\usepackage[dvips]{graphicx}
\usepackage{pstricks,pst-coil}
\usepackage{pst-all}
\RequirePackage{flushend}
\RequirePackage[colorlinks,citecolor=blue,urlcolor=blue,linkcolor=blue]{hyperref}
\usepackage{color}
\usepackage{latexsym}
\usepackage{amsmath}
\usepackage{amssymb}
\usepackage{eufrak}
\usepackage{euscript}
\usepackage{graphics}
\usepackage{picture}
\def\[{\left\lbrack}
\def\]{\right\rbrack}

\def\({\left(}
\def\){\right)}

\newcommand{\bbe}{\begin{equation}}
\newcommand{\eee}{\end{equation}}
\newcommand{\eaa}{\end{eqnarray}}
\newcommand{\baa}{\begin{eqnarray}}

\def\uma{\rm 1\!\!\hskip 1 pt l}

\begin{document}


\title{\Large{A $Z'$-model and the magnetism of a dark fermion candidate}}

\author{M. J. Neves}
\email{mariojr@ufrrj.br}
\affiliation{Departamento de F\'{\i}sica, Universidade Federal Rural do Rio de Janeiro,
BR 465-07, 23890-971, Serop\'edica, RJ, Brazil}

\author{J. A. Helay\"el-Neto}
\email{helayel@cbpf.br}
\affiliation{Centro Brasileiro de Pesquisas F\' isicas (CBPF),
\\
Rua Dr. Xavier Sigaud 150, Urca,
CEP  22290-180, Rio de Janeiro, Brazil}

\date{\today}


\begin{abstract}
\noindent

Our contribution sets out to investigate the phenomenology of a gauge model based on an $SU_{L}(2) \times U_{R}(1)_{Q} \times U(1)_{Q'}$-symmetry group. The model can accommodate, through its symmetry-breaking pattern, a candidate to a heavy $Z'$-boson at the TeV-scale. The extended Higgs sector introduces a heavy scalar whose mass lies in the region $1.2-3.7 \, \mbox{TeV}$. The fermion sector includes an exotic candidate to Dark Matter that mixes with the right-handed neutrino component in the Higgs sector, so that the whole field content ensures the cancellation of the $U(1)$-quiral anomaly.
The masses are fixed according to the particular way the symmetry breaking takes place. In view of the possible symmetry breakdown pattern, we study the phenomenological implications in a high-energy scenario. We worek out the magnetic dipole momentum (MDM) of the exotic fermion and the transition MDM due to its mixing with the right-neutrino.
\end{abstract}
%
%
%
%

\maketitle

\pagestyle{myheadings}
\markright{The magnetic dipole momentum of Dark fermion candidate out of a $Z'$-model}

%
%
%

%
\section{Introduction}
\renewcommand{\theequation}{1.\arabic{equation}}
\setcounter{equation}{0}

The search for new particles and interactions beyond the Standard Model (SM) has been challenging High-Energy Physics, both Theoretical and
Experimental, over the recent decades. The results from the LHC's ATLAS- and CMS-Collaborations can point to the existence of a (new) fifth
interaction at the TeV-scale and beyond \cite{Atlas,CMS}. The search for heavy resonances in proton-proton collisions at a center-of-mass energy scale of $\sqrt{s}= 8 \, \mbox{TeV}$ excludes $Z'$ masses below the $2.5 \, \mbox{TeV}$ in the decay process of $Z'$ into leptonic final states \cite{CMS20171}. This result indicates that other cascade effects might reveal new resonances for collision energy scales above $10 \, \mbox{TeV}$.
For example, the $Z'$-cascade effects can connect the Standard Model matter sector with possible Dark Matter (DM) fermionic particles.
Several simulations with vector, axial and pseudo-scalar mediators are discussed in mass range between
$0.15 \, \mbox{TeV}$ and $2.5 \, \mbox{TeV}$ for $Z'$, and $50 \, \mbox{GeV}$ to $1.2 \, \mbox{TeV}$ for the DM sector; for that, we refer to \cite{CMS20172}.


%

%
A well-known model in this direction is based on an $SU_{L}(2)\times SU_{R}(2) \times U(1)_{B-L}$-gauge symmetry. One refers to this
proposal as Left-Right Symmetric Model \cite{DobrescuJHEP2015,DobrescuPRL2015,DobrescuPRD2015,DobrescuJHEP2016,DevPRL2015,Patra2016,HongGuPRD2017,Dev2017}.
The extra $SU_{R}(2)$-subgroup is introduced to account for the $W^{\, \prime}$- and $Z^{\, \prime}$-resonances at the TeV-scale,
besides the already-known $W^{\pm}$- and $Z^{0}$-weak mediators of the Glashow-Salam-Weinberg (GSW) model.
The right-handed sector also introduces fermion doublets with a charged lepton and its associated neutrino of right chirality.
Three scalar triplets constitute the Higgs sector responsible for the chain of symmetry breakings.
The triplets acquire three vacuum expected values (VEV) scales that fix the $W^{\, \prime}$ and $Z^{\, \prime}$ masses at the TeV scale,
the usual $W$ and $Z$ masses in the SM at the GeV scale and the light and heavy neutrinos masses via a see-saw mechanism.
%

%
%
%
%
%

The well-known model in the literature that describes only the $Z'$-heavy boson is based on the gauge symmetry $SU_{L}(2) \times
U_{Y}(1) \times U(1)_{B-L}$ \cite{LangackerRMP2009,Kanemura2011,MorettiarXiv2017}, which includes an extra $U(1)$-group in the GSW model.
The gauge sector has just one extra boson, while the Higgs is extended to include a doublet and a singlet, or bi-doublets of scalars fields.
We should keep in mind that the introduction of an extra Higgs may be needed to explain the heavy masses of the new bosons at the $\mbox{TeV}$-scale.
These masses are associated with a new range of VEVs of the extra Higgs scalars. The fermion sector is enlarged to guarantee anomaly cancellation; for example, the introduction of right-neutrino components and exotic fermion singlets that could be candidates to a DM sector.

In the present paper, we re-assess the $SU_{L}(2) \times U_{R}(1)_{Q} \times U(1)_{Q'}$ model whose
SSB pattern can be contemplated in a higher energy scale; for details, consult \cite{MJNeves2017Annalen}.
We study its phenomenology such that the $Z'$-boson can connect the world of SM particles with exotic fermions
candidate to DM particles through the cascade effects. Furthermore, we introduce a mixing sub-sector involving a right-neutrino
component and exotic fermions through Yukawa interactions in the extended Higgs sector,
{\it i. e.}, heavy and neutral fermions can unveil relevant magnetic properties of these new fermions. We pursue this investigation
and we obtain the Transition Magnetic Dipole Momentum (TMDM) and the Magnetic Dipole Momentum (MDM)
for the exotic fermion in terms of its mass, as it happens in the neutrino case. Sectors of fermions and scalar bosons are introduced with quantum numbers consistent with the gauge invariance so that the chiral anomalies associated with the Abelian sectors cancel out. The mixing of the SM neutrinos with an exotic DM-candidate fermion and the detailed discussion on their the MDMs are two other aspects that we wish to point in our endeavour.

The organization of this paper follows the outline below: in Section II, we review the $Z'$-model based on an $SU_{L}(2) \times U_{R}(1)_{Q} \times U(1)_{Q'}$-symmetry presented in details in \cite{MJNeves2017Annalen}. In Section III, we obtain the $Z'$-mass and it coupling with the fermions of the SM. Section IV deals with the details of the diagonalization of the mass matrix of the dark fermions and right-neutrinos in the extended Higgs sector. Section V is presented in three subsections, where we study the $Z'$-decay modes into fermions and the decay modes of the extra Higgs .
In Section VI, we obtain the TMDM and the MDM of the dark fermions. Finally, our Concluding Comments are cast in Section VII.
%
%
%

\section{A glance at the $Z'$-model}
\renewcommand{\theequation}{2.\arabic{equation}}
\setcounter{equation}{0}

In this Section, we present a short review of the $Z^{\prime}$-model; the details may be found in the work of Ref. \cite{MJNeves2017Annalen}.
Here, we introduce the model in the renormalizable $R_{\xi}$-gauge. The sector of fermions and gauge fields of the model
$SU_{L}(2) \times U_{R}(1)_{Q} \times U(1)_{Q'}$-model is described by the Lagrangian below :
\begin{eqnarray}\label{Lleptons}
{\cal L}_{f} \!\!&=&\!\! \bar{\Psi}_{L}\, i \, \, \slash{\!\!\!\!D} \, \Psi_{L}
+\bar{\Psi}_{R} \, i \, \, \slash{\!\!\!\!D} \, \Psi_{R}
+\bar{\nu}_{iR} \, i \, \, \slash{\!\!\!\!D} \, \nu_{iR}
\nonumber \\
&&
+\bar{\zeta}_{iL} \, i \, \, \slash{\!\!\!\!D} \, \zeta_{iL}+\bar{\zeta}_{iR} \, i \, \, \slash{\!\!\!\!D} \, \zeta_{iR}  \; ,
\end{eqnarray}
and
\begin{equation}\label{Lgauge}
{\cal L}_{gauge}=-\frac{1}{2} \,\mbox{tr}\left(F_{\mu\nu}^{\; 2}\right)
-\frac{1}{4} \, B_{\mu\nu}^{\; 2}
-\frac{1}{4} \, C_{\mu\nu}^{\; 2} \; .
\end{equation}
The slashed notation corresponds to the contraction of the covariant derivatives with the usual Dirac matrices.
The covariant derivatives acting on the fermions of the model are collected below:
\begin{eqnarray}\label{DmuPsiLPsiR}
D_{\mu}\Psi_{L} \!\!&=&\!\! \left(\partial_{\mu}+i \, g \, A_{\mu}^{\, a} \, \frac{\sigma^{a}}{2} +i \, Q_{L} \, g^{\prime} \, B_{\mu}
+ i \, Q'_{L} \, g^{\prime \prime} \, C_{\mu} \! \right) \! \Psi_{L} \; ,
\nonumber \\
D_{\mu}\Psi_{R}\!\!&=&\!\! \left( \phantom{\frac{1}{2}} \!\!\!\! \partial_{\mu} + i \, Q_{R} \, g^{\prime} \, B_{\mu}+i \, Q'_{R} \, g^{\prime \prime} \, C_{\mu} \right) \Psi_{R} \; ,
\nonumber \\
D_{\mu}\nu_{iR} \!\!&=&\!\! \left( \phantom{\frac{1}{2}} \!\!\!\! \partial_{\mu} + i \, Q_{\nu_{R}} \, g^{\prime} \, B_{\mu}+i \, Q'_{\nu_{R}} \, g^{\prime \prime} \, C_{\mu} \right) \nu_{iR} \; ,
\nonumber \\
D_{\mu}\zeta_{iL} \!\!&=&\!\! \left( \phantom{\frac{1}{2}} \!\!\!\! \partial_{\mu}+ i \, Q_{\zeta_{L}} \, g^{\prime} \, B_{\mu} + i \, Q'_{\zeta_{L}} \, g^{\prime \prime} \, C_{\mu} \right) \zeta_{iL} \; ,
\nonumber \\
D_{\mu}\zeta_{iR} \!\!&=&\!\! \left( \phantom{\frac{1}{2}} \!\!\!\! \partial_{\mu}+ i \, Q_{\zeta_{R}} \, g^{\prime} \, B_{\mu} + i \, Q'_{\zeta_{R}} \, g^{\prime \prime} \, C_{\mu} \right) \zeta_{iR}
\; ,
\end{eqnarray}
where $A^{\mu \, a}\!=\!\left\{ \, A^{\mu \, 1} \, , \, A^{\mu \, 2} \, , \, A^{\mu \, 3} \, \right\}$ are the gauge fields of $SU_{L}(2)$, $B^{\mu}$ is the Abelian gauge field of $U_{R}(1)_{Q}$, and $C^{\mu}$ the similar one to $U(1)_{Q'}$. Here, we have chosen the symbol $J$ to represent the generator of $U_{R}(1)_{Q}$, $Q'$ is the generator of $U(1)_{Q'}$, the generators of $SU_{L}(2)$ are the Pauli matrices $\frac{\sigma^{a}}{2} \, (a=1,2,3)$, and
$g$, $g^{\, \prime}$ and $g^{\, \prime\prime}$ are dimensionless gauge couplings. The fermionic field content is given in the sequel.
The notation $\Psi_{L}$ indicates the usual doublet of neutrinos/leptons $L_{i}=(\nu_{i} \, \, \, \ell_{i})_{L}^{t}$,
or quarks $Q_{iL}$ $(i=1,2,3)$ left-handed of the SM, which transform under the fundamental representation of $SU_{L}(2)$.
The label $\ell_{iL}=\left( \, e_{L} \, , \, \mu_{L} \, , \, \tau_{L} \, \right)$ indicates the leptons family displayed in the doublet, and
the $\nu_{iL}$-label for neutrinos is defined by $\nu_{iL}=(\nu_{e_{L}},\nu_{\mu_{L}},\nu_{\tau_{L}})$.
The $\Psi_{R}$-fermion is any right-handed field of the SM, {\it i. e.}, it may be the lepton
$\ell_{iR}=\left( \, e_{R} \, , \, \mu_{R} \, , \, \tau_{R} \, \right)$ or the right-handed quarks
$Q_{iR}=\left\{ \, u_{R} \, , \, c_{R} \, , \, t_{R} \, \right\}$ and $q_{iR}=\left\{ \, d_{R} \, , \, s_{R} \, , \, b_{R} \, \right\}$.
The new content of fermions beyond the SM is constitued by the Dirac Right-Neutrino $\nu_{iR}=(\nu_{e_{R}},\nu_{\mu_{R}},\nu_{\tau_{R}})$
and the exotic neutral $\zeta_{iL(R)}$-fermion $(i=1,2,3)$ that we have introduced in association with the $U(1)_{Q'}$-group:
\begin{eqnarray}
L_{i}&=&
\left(
\begin{array}{c}
\nu_{e} \\
\ell_{e} \\
\end{array}
\right)_{L} ,
\left(
\begin{array}{c}
\nu_{\mu} \\
\ell_{\mu} \\
\end{array}
\right)_{L} ,
\left(
\begin{array}{c}
\nu_{\tau} \\
\ell_{\tau} \\
\end{array}
\right)_{L}
\hspace{-0.1cm}
: \left(\underline{{\bf 2}}, 0, -\frac{1}{2} \right) \, ,
\nonumber \\
Q_{iL}&=&
\left(
\begin{array}{c}
u \\
d \\
\end{array}
\right)_{L} ,
\left(
\begin{array}{c}
c \\
s \\
\end{array}
\right)_{L} ,
\left(
\begin{array}{c}
t \\
b \\
\end{array}
\right)_{L}
\hspace{-0.1cm}
: \left(\underline{{\bf 2}}, 0, +\frac{1}{6} \right) \, ,
\nonumber
\end{eqnarray}
\vspace{-0.5cm}
\begin{eqnarray}
\ell_{iR} &=& \left\{ \, e_{R} \, , \, \mu_{R} \, , \, \tau_{R} \, \right\} : \left(\underline{{\bf 1}}, -\frac{1}{2}, -\frac{1}{2} \right) \, ,
\nonumber \\
Q_{iR} &=& \left\{ \, u_{R} \, , \, c_{R} \, , \, t_{R} \, \right\} : \left(\underline{{\bf 1}}, +\frac{1}{2}, +\frac{1}{6} \right) \, ,
\nonumber \\
q_{iR} &=& \left\{ \, d_{R} \, , \, s_{R} \, , \, b_{R} \, \right\} : \left(\underline{{\bf 1}}, -\frac{1}{2}, +\frac{1}{6} \right) \, ,
\nonumber \\
\nu_{iR} &=& \left\{ \, \nu_{eR} \, , \, \nu_{\mu R} \, , \, \nu_{\tau R} \, \right\} : \left(\underline{{\bf 1}}, +\frac{1}{2}, -\frac{1}{2} \right) \, ,
\nonumber \\
\zeta_{iL} &=& \left\{ \, \zeta_{1L} \, , \, \zeta_{2L} \, , \, \zeta_{3L} \, \right\} : \left(\underline{{\bf 1}}, +\frac{1}{2}, -\frac{1}{2} \right) \, ,
\nonumber \\
\zeta_{iR} &=& \left\{ \, \zeta_{1R} \, , \, \zeta_{2R} \, , \, \zeta_{3R}  \, \right\} : \left(\underline{{\bf 1}}, -\frac{1}{2}, +\frac{1}{2} \right) \, .
\end{eqnarray}
All these fields are singlets that undergo Abelian transformations under the $U_{R}(1)_{J}$- and $U(1)_{K}$-groups.
The corresponding field-strength tensors in the sector of gauge fields are defined by
\begin{eqnarray}\label{Fmunu}
F_{\mu\nu} \!\!&=&\!\! \partial_{\mu}A_{\nu}-\partial_{\nu}A_{\mu}+i \, g \, \left[ \, A_{\mu} \, , \, A_{\nu} \, \right]
\; , \;
\nonumber \\
B_{\mu\nu} \!\!&=&\!\! \partial_{\mu}B_{\nu}-\partial_{\nu}B_{\mu}
\hspace{0.2cm} \mbox{and} \hspace{0.2cm}
C_{\mu\nu}= \partial_{\mu}C_{\nu}-\partial_{\nu}C_{\mu} \; .
\hspace{0.5cm}
\end{eqnarray}
%
%
%

%

%

%
%
%
%
%
%
%
%

The Higgs sector is essential to introduce the masses,
the physical fields and the charges for the particle content
of the model. The content of the Higgs sector is given by
two independent scalar fields; the first is a singlet
scalar, $\Xi$, that breaks the Abelian subgroup to generate
mass to the new gauge boson that, in this scenario, we refer to
as $Z'$-boson. The second Higgs field, $\Phi$, is an $SU_{L}(2)$-
doublet that breaks the residual electroweak symmetry and,
consequently, yields the known masses for $W^{\pm}$ and
$Z^{0}$. Finally, we end up with the exact electromagnetic
symmetry
\begin{equation}
SU_{L}(2) \times U_{R}(1)_{Q} \times U(1)_{Q'} \stackrel{\langle \Xi \rangle_{0}}{\longmapsto}
SU_{L}(2) \times U_{Y}(1) \stackrel{\langle \Phi \rangle_{0}}{\longmapsto} U_{em}(1) \; ,
\end{equation}
where the $U_{Y}(1)$-group comes out as the mixing of the $U_{R}(1)_{Q}$- and $U(1)_{Q'}$-subgroups.
To accomplish this SSB pattern, we start off from the Higgs Lagrangian below:
\begin{eqnarray}\label{LHiggs}
{\cal L}_{Higgs} \!\!&=&\!\!
\left(D_{\mu}\Xi\right)^{\dagger} D^{\mu} \Xi
-\mu_{\Xi}^{\, 2} \, |\Xi|^{2} -\lambda_{\Xi} \, |\Xi|^{4}
\nonumber \\
&&
\hspace{-0.7cm}
+\left(D_{\mu}\Phi\right)^{\dagger} D^{\mu} \Phi
-\mu_{\Phi}^{\, 2} \, |\Phi|^{2}-\lambda_{\Phi} \, |\Phi|^{4}
-\lambda \, |\Xi|^{\, 2} \, |\Phi|^{\, 2} \, ,
\nonumber \\
&&
\hspace{-0.7cm}
- \, G_{ij}^{(\ell)} \, \bar{L}_{i} \, \Phi \, \ell_{jR}
- \, G_{ij}^{(d)} \, \bar{Q}_{iL} \, \Phi \, q_{jR}
- \, G_{ij}^{(u)} \, \bar{Q}_{iL} \, \widetilde{\Phi} \, Q_{jR}
\nonumber \\
&&
\hspace{-0.7cm}
- \, X_{ij} \, \bar{L}_{i} \, \widetilde{\Phi} \, \nu_{j R}
- \, Y_{ij} \, \bar{L}_{i} \, \widetilde{\Phi} \, \zeta_{jR}
\nonumber \\
&&
\hspace{-0.5cm}
- \, Z_{ij} \, \bar{\zeta}_{iL} \, \Xi \, \nu_{j R}
- \, W_{ij} \, \bar{\zeta}_{iL} \, \Xi \, \zeta_{jR}+\mbox{h. c. }
\; .
\end{eqnarray}
In (\ref{LHiggs}), $\left\{ \, \mu_{\Xi} \, , \, \mu_{\Phi} \, , \, \lambda_{\Xi} \, , \, \lambda_{\Phi} \, , \, \lambda\, \right\}$ are real parameters,
$\left\{ \, G_{ij}^{(\ell)} \, , \, G_{ij}^{(u)} \, , \, G_{ij}^{(d)} \, , \, X_{ij} \, , \, Y_{ij} \, , \, Z_{ij} \, , \, W_{ij} \, \right\}$
are Yukawa (complex) coupling parameters needed for the fermions to acquire non-trivial masses. In general, these Yukawa couplings
set non-diagonal matrices $3 \times 3$,
and as usual, the $\widetilde{\Phi}$-field is defined as $\widetilde{\Phi}=i \, \sigma_{2} \, \Phi^{\ast}$ to ensure the gauge invariance.

The covariant derivatives of (\ref{LHiggs}) act on the $\Xi$- and $\Phi$-Higgs as follows:
\begin{eqnarray}\label{DmuPhi1}
D_{\mu} \, \Xi(x) \!\!&=&\!\! \left( \phantom{\frac{1}{2}} \!\!\!\! \partial_{\mu}+i \, Q_{\Xi} \, g^{\prime} B_{\mu}+i \, Q'_{\Xi} \, g^{\prime\prime}
C_{\mu} \right)\Xi(x)
\nonumber \\
D_{\mu} \Phi(x) \!\!&=&\!\! \left(\partial_{\mu}
+ i g \, A_{\mu}^{\, a} \, \frac{\sigma^{a}}{2}+ i g'  \, Q_{\Phi} \, B_{\mu} \right) \Phi(x) \; ,
\; \;
\end{eqnarray}
where the $Q_{\Phi}$ is the generator of $\Phi$-Higgs corresponding to the $U_{R}(1)_{Q}$-subgroup.
The $\Xi$-field is a scalar singlet of $SU_{L}(2)$, with transformations under $U_{R}(1)_{Q} \times U(1)_{Q'}$.
%
%
%
The second $\Phi$-scalar field is a doublet $\Phi=\left( \, \phi^{ \, (+)} \; \;
\phi^{\, (0)} \, \right)^{t}$ that turns out in the fundamental representation of $SU_{L}(2)$, and it also transforms under $U_{R}(1)_{Q}$ subgroup.   
%
%
%
%
%
%
%
We choose the parametrization of the $\Xi$- and $\Phi$-complex fields as
\begin{eqnarray}\label{PhiGaugeparametrization}
\Xi(x) \!\!&=&\!\! \frac{u+F(x)}{\sqrt{2}} \, \, e^{\, i \, \frac{\eta(x)}{u}}
\hspace{0.2cm} , \hspace{0.2cm}
\nonumber \\
\Phi(x) \!\!&=&\!\! \exp\left[\, \frac{i}{v}
\left(
\begin{array}{cc}
\chi^{3} & \sqrt{2} \, \chi^{-} \\
\sqrt{2} \, \chi^{+} & -\chi^{3} \\
\end{array}
\right)
\right]
\left(
\begin{array}{c}
0 \\
\frac{v+H(x)}{\sqrt{2}} \\
\end{array}
\right) \, ,
\hspace{0.5cm}
\end{eqnarray}
where ˜$F$, $˜H$ are real functions, and $\{ \, \eta \, , \, \chi^{\pm} \, , \, \chi_{3} \, \}$ are the four would-be-Goldstone
bosons, such that the charged Goldstone bosons are defined by $\sqrt{2} \, \chi^{\pm}:=\chi^{1} \, \mp \, i \, \chi^{2}$.
The minima of the Higgs potential are given by the non-trivial VEVs $\{ \, u \, , \, v \, \}$ listed below:
\begin{eqnarray}
\langle \Xi \rangle_{0} \!\!&=&\!\! \frac{u}{\sqrt{2}} \simeq \sqrt{-\frac{\mu_{\Xi}^{2}}{2\lambda_{\Xi}}} \left(1- \frac{\lambda}{4 \, \lambda_{\Phi}}
\frac{\mu_{\Phi}^{2}}{\mu_{\Xi}^{2}}  \right) \, ,
\nonumber \\
\langle \Phi \rangle_{0} \!\!&=&\!\! \frac{v}{\sqrt{2}} \simeq \sqrt{-\frac{\mu_{\Phi}^{2}}{2\lambda_{\Phi}}}  \left( 1- \frac{\lambda}{4 \, \lambda_{\Xi}}
\frac{\mu_{\Xi}^{2}}{\mu_{\Phi}^{2}} \right) \, ,
\end{eqnarray}
where the following conditions are satisfied : $\mu_{\Xi}^{2}<0$ , $\mu_{\Phi}^{2}<0$ and $\lambda \ll 1$.
It is important to emphasize that, in the
$Z^{\prime}$-approach, the necessary condition $u \gg v$ between the
VEVs must be satisfied, such that $u$-scale generates mass
for the heavy $Z^{\prime}$-boson, while the $v = 246 \, \mbox{GeV}$ is the well-known electroweak scale of the SM.

\section{The mass eigenstates of Z and Z'}

\begin{figure}[h]
\centering
\includegraphics[scale=0.47]{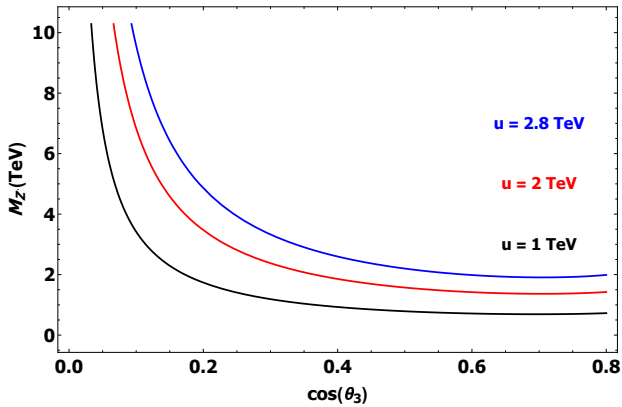}
\caption{The $Z'$-mass as function of the $\theta_{3}$-mixing angle for some values of $u$-VEV scale.
The blue line indicates the case of $u=2.8 \, \mbox{TeV}$, when $\theta_{3}=45^{0}$, the $Z'$-mass is around $2 \, \mbox{TeV}$.
The other cases correspond to $u=2 \, \mbox{TeV}$ (red line), and for $u=1 \, \mbox{TeV}$ (black line).}
\label{MZ'theta3}
\end{figure}

After the SSBs, the gauge sector apperas as
\begin{eqnarray}\label{massesBWZ}
{\cal L}_{mass} \!\!&=&\!\!
m_{W}^{\, 2} \, W_{\mu}^{+}W^{\mu-}+
\frac{u^2}{2} \left( \phantom{\frac{1}{2}} \!\!\!\!\! g^{\prime} B_{\mu}- \, g^{\prime \prime} C_{\mu} \right)^2
\nonumber \\
&&
\hspace{-0.4cm}
+\frac{v^{2}}{8} \left( \phantom{\frac{1}{2}} \hspace{-0.3cm} g^{\prime} \, B_{\mu}- \, g \, A_{\mu}^{3} \right)^{2}
 .
\hspace{0.5cm}
\end{eqnarray}
The $W^{\pm}$-particles are identified as the combination $\sqrt{2} \, W_{\mu}^{\pm}:=A_{\mu}^{1} \, \mp \, i \, A_{\mu}^{2}$,
and it $W^{\pm}$- mass is like in GSW-model $M_{W}= gv/2$. The neutral sector of (\ref{massesBWZ}) can be cast into a
matrix form
\begin{eqnarray}
{\cal L}_{mass}=V_{\mu}^{\, t} \, \eta^{\mu\nu} \, M_{Z-Z'}^{\, 2} V_{\nu} \; ,
\end{eqnarray}
where we have defined the column-vector
$V^{\mu \, t}=\left( \, A^{\mu3} \; \; B^{\mu} \; \; C^{\mu} \, \right)$, and the mass matrix $M_{Z-Z^{\prime}}^{2}$
is given by
\begin{eqnarray}
M_{Z-Z'}^{\, 2}\!=\!
\left(
\begin{array}{ccc}
g^{2}v^{2}/4 & -gg^{\prime}v^2/4 & 0 \\
-gg^{\prime}v^2/4 & g^{\prime \, 2} (u^2+v^2/4) & -g^{\prime}g^{\prime\prime}u^2 \\
0 & -g^{\prime}g^{\prime\prime}u^2 & g^{\prime\prime \, 2} u^2 \\
\end{array}
\right) .
\hspace{0.4cm}
\end{eqnarray}
The mass matrix is diagonalized by the orthogonal $SO(3)$-transformation $V^{\prime \, \mu}=R \, V^{\mu}$,
where $R$ is the most general $SO(3)$-matrix, {\it i. e.}, $R=R_{1}(\theta_{1}) \, R_{2}(\theta_{2}) \, R_{3}(\theta_{3})$,
with the mixing angles $\theta_{1}$, $\theta_{2}$ and $\theta_{3}$, and $R_{i}$ are special rotation matrices. In the $V^{\prime \, \mu}$-basis,
the diagonal mass matrix is given by
\begin{eqnarray}\label{MD}
M_{D}^{2}=R \, M^{2} \, R^{t}=
\left(
\begin{array}{ccc}
0 & 0 & 0 \\
0 & M_{Z'}^{\, 2} & 0 \\
0 & 0 & M_{Z}^{\, 2} \\
\end{array}
\right) \; ,
\end{eqnarray}
whose the eigenvalues, for the condition $u \gg v$, are read below
\begin{eqnarray}
M_{Z'} \!\!&\simeq&\!\! u \sqrt{g^{\prime \, 2}+g^{\prime\prime \, 2}}\left[1+\frac{v^2}{8u^2} \frac{g^{\prime \, 4}}{(g^{\prime \, 2}+g^{\prime\prime \, 2})^{2}} \right] \; ,
\nonumber \\
M_{Z} \!\! &\simeq& \!\! \frac{v}{2} \, \sqrt{\frac{g^{2}g^{\prime \, 2}+g^{2}g^{\prime\prime \, 2}+g^{\prime \, 2}g^{\prime\prime \, 2}}{g^{\prime \, 2}+g^{\prime\prime \, 2}}} \; .
\end{eqnarray}
The null eigenvalue of (\ref{MD}) is identified as the photon mass, thus,
we can denote the column-vector in the mass basis as
$V^{\prime \, \mu \, t}=\left( \, A^{\mu} \; \; Z^{\prime \, \mu} \; \; Z^{\mu} \, \right)$.
Therefore, we obtain the basis transformation :
\begin{eqnarray}\label{transfA0CGY}
C_{\mu} \!\!&=&\!\! \cos\theta_{2}\cos\theta_{3} \, A_{\mu}
\nonumber \\
&&
\hspace{-0.5cm}
- \left(\cos\theta_{3}\sin\theta_{1} \sin\theta_{2}+ \cos\theta_{1}\sin\theta_{3} \right) Z'_{\mu}
\nonumber \\
&&
\hspace{-0.5cm}
+\left(\sin\theta_{1}\sin\theta_{3}-\cos\theta_{1}\cos\theta_{3}\sin\theta_{2}\right) Z_{\mu} \; ,
\nonumber \\
B_{\mu} \!\! &=& \!\! \sin\theta_{3} \cos\theta_{2} \, A_{\mu}
\nonumber \\
&&
\hspace{-0.5cm}
+ \left(\cos\theta_{1}\cos\theta_{3}-\sin\theta_{1}\sin\theta_{2}\sin\theta_{3} \right) Z'_{\mu}
\nonumber \\
&&
\hspace{-0.5cm}
-\left(\sin\theta_{1}\cos\theta_{3}+\cos\theta_{1}\sin\theta_{2}\sin\theta_{3} \right) Z_{\mu} \; ,
\nonumber \\
A_{\mu}^{\, 3} \!\!&=&\!\! \sin\theta_{2} \, A_{\mu}+ \sin\theta_{1} \cos\theta_{2} \, Z'_{\mu}+ \cos\theta_{1} \cos\theta_{2} \, Z_{\mu} \; ,
\hspace{1cm}
\end{eqnarray}
where we choose the parametrization for the fundamental charge $e^2=4\pi/137\simeq 0.09$
\begin{eqnarray}\label{gYgg}
e=g \sin\theta_{2}=g^{\prime}\cos\theta_{2} \sin\theta_{3} = g^{\prime\prime} \cos\theta_{2}\cos\theta_{3} \; .
\end{eqnarray}
This parametrization indicates that $\theta_{2}$ is the Weinberg
angle $\sin^2 \theta_{2}=\sin^2 \theta_{W} = 0.22$
\footnote{We are considering the value of the Weinberg angle as
$\sin^{2}\theta_{W} = 0.22$, taking into account the radiative corrections of the Electroweak Theory.}.
Therefore, using the parametrization (\ref{gYgg}), the masses of the $W$- $Z$-bosons are those known in the GSW model :
$M_{W}=78 \, \mbox{GeV}$ and $M_{Z}= 90 \, \mbox{GeV}$ at the tree level, in which the experimental data have
been taken into \cite{PDG2016}. The $Z'$-mass in terms of the fundamental
charge, $u$-VEV scale, and the $\theta_{3}$ mixing angle is given by
\begin{eqnarray}\label{massesWZZ'}
M_{Z'} \!\! &\simeq& \!\! \frac{0.68 \, u}{\sin(2\theta_{3})} \! \left( 1 + \frac{v^{2}}{8 \, u^{2}} \, \cos^{4}\theta_{3} \right) \; .
\end{eqnarray}
The $u$-scale and the $\theta_{3}$-angle heve not been determined in the previous expression for the $Z'$-mass.
We illustrate a plot of the $Z'$-mass of (\ref{massesWZZ'}) as a function of the $\theta_{3}$-mixing angle; this is depicted in figure (\ref{MZ'theta3}).
The ratio between the masses of $Z$ and $Z'$ from (\ref{massesWZZ'}) is given by
\begin{eqnarray}
\frac{M_{Z'}}{M_{Z}} \simeq \frac{u}{\sin(2\theta_{3})} \, \frac{4\sin\theta_{W}}{v} \simeq \frac{7u}{\sin(2\theta_{3})} \, \mbox{TeV}^{-1} .
\hspace{0.5cm}
\end{eqnarray}
The recent papers of the CMS Collaboration point to the hypothetical $Z'$ upper limits
that exclude up to $95\%$ confidence level masses below the $2.0 \, \mbox{TeV}$ \cite{CMS20172}.
Therefore, we fix the $Z$- and $Z'$-masses as $M_{Z'}=2.0 \, \mbox{TeV}$ and $M_{Z}=91 \, \mbox{GeV}$
to estimate the ratio of the $u$-scale by the $\theta_{3}$-angle, {\it i. e.}, $u \simeq 2.8 \, \sin(2\theta_{3}) \, \mbox{TeV}$.
In so doing, the maximum value for the VEV-scale is $u=2.8 \, \mbox{TeV}$, when $\theta_{3}=45^{o}$ and $M_{Z'}=2.0 \, \mbox{TeV}$.
This case is shown in figure (\ref{MZ'theta3}) (blue line). The case $u= 2 \, \mbox{TeV}$ (red line) fixes a lower bound
of $2 \, \mbox{TeV}$ for the $Z'$-mass, while $u = 1 \, \mbox{TeV}$ (black line) fixes a lower bound below $2 \, \mbox{TeV}$
for the $Z'$-mass in the range $0.4< \cos\theta_{3} <0.8$. The Higgs sector, after the SSBs, is reduced to the scalar fields ˜$F$ and $H$, whose the mass matrix can be read below :
\begin{eqnarray}\label{MFH}
M_{F-H}^{\, 2}=\!
\left( \!\!
\begin{array}{cc}
m_{F}^{\, 2}  & - \, \frac{\tilde{\lambda}}{2} \, m_{H} \, m_{F}
\\
\\
-\, \frac{\tilde{\lambda}}{2} \, m_{H} \, m_{F} & m_{H}^{\, 2}   \\
\end{array}
\!\!\right) . \;\;
\end{eqnarray}
Here, $m_{F}=\sqrt{2 \lambda_{\Xi} \, u^{2}}$ and $m_{H}=\sqrt{2 \lambda_{\Phi} \, v^{2}}$ are the masses of $H$- and $H$-fields,
when $\lambda \rightarrow 0$, and $\tilde{\lambda}:=\lambda/\sqrt{\lambda_{\Phi}\lambda_{\Xi}}$ for short. Orthogonal transformations
yield the combinations that correspond to the physical fields, $F'$ and $H'$ (mass basis for $F$ and $H$)
\begin{eqnarray}
\left(
\begin{array}{c}
F' \\
\\
H' \\
\end{array}
\right)
=\left(
\begin{array}{cc}
\cos\vartheta & \sin\vartheta
\\
\\
-\sin\vartheta & \cos\vartheta \\
\end{array}
\right)
\left(
\begin{array}{c}
F \\
\\
H \\
\end{array}
\right) \; ,
\end{eqnarray}
where $\vartheta$ is the mixing angle defined by
\begin{eqnarray}\label{thetaZZm}
\tan2\vartheta=-\frac{\tilde{\lambda} \, m_{H} \, m_{F} }{m_{F}^{\, 2}-m_{H}^{\, 2}}\simeq
- \tilde{\lambda} \; \frac{m_{H}}{m_{F}} \; .
\end{eqnarray}
From that, the eigenvalues of the mass matrix (\ref{MFH}) yield the physical masses
\begin{eqnarray}\label{MFH}
M_{H} \!\!&\simeq&\!\! \sqrt{ 2 \, \lambda_{\Phi} \, v^{2}}
\, \left( 1-\frac{\lambda^2}{8 \, \lambda_{\Xi}\lambda_{\Phi}} \right)=125 \, \mbox{GeV}
\nonumber \\
M_{F} \!\!&\simeq&\!\! \sqrt{ 2 \, \lambda_{\Xi} \, u^{2}}
\, \left( 1+\frac{\lambda^2}{8 \, \lambda_{\Xi}^2} \, \frac{v^{2}}{u^{2}}  \right)
\; ,
\end{eqnarray}
%
%
%
%
where $M_{F} \gg M_{H}$ as a consequence of the condition on the VEV-scales. The estimation for $M_{F}$ is
\begin{eqnarray}\label{MassH1u8GEV}
1.24 \, \mbox{TeV} < M_{F} < 3.7 \, \mbox{TeV} \; ,
\end{eqnarray}
in which the maximum value of $3.7 \, \mbox{TeV}$ correspond to $\theta_{3}$-angle of $45^{o}$.

The interaction vertices of the $Z$-, $Z'$-bosons and the photon
with the fermions of the model follow below:
\begin{eqnarray}\label{LintAZC}
{\cal L}^{\, int} \!=\! - \, e \, \bar{\Psi}_{L(R)} \left( \, Q_{em} \, \, \slash{\!\!\!\!A}
+ Q_{Z} \, \slash{\!\!\!\!Z} +Q_{Z'} \, \slash{\!\!\!\!Z'} \, \right) \Psi_{L(R)} \; .
\hspace{0.5cm}
\end{eqnarray}
The electric charge operator of the model is
\begin{eqnarray}
Q_{em}=I^{3}+Q+Q' \; ,
\end{eqnarray}
and the generators $Q_{Z}$ and $Q_{Z'}$ are defined by the relation
\begin{eqnarray}\label{Nishijima}
Q_{Z} \!&=&\! \frac{I^{3}- Q_{em} \, \sin^{2}\theta_{W} }{\sin\theta_{W}\cos\theta_{W}} +
\nonumber \\
&&
\hspace{-0.8cm}
+ \sin\theta_{1}\sec\theta_{W}\left( \tan\theta_{3}Q'-\cot\theta_{3}Q \right) \; ,
\nonumber \\
Q_{Z'} \!&=&\! \frac{2 \cos\theta_{1}}{\sin(2\theta_{3}) \cos\theta_{W}} \left( \phantom{\frac{1}{2}} \!\!\!\!\!\! - Q \, + Y \, \sin^{2}\theta_{3} \right)
\nonumber \\
&&
\hspace{-0.8cm}
+\sin\theta_{1}\left( \cot\theta_{3} I^{3}-\tan\theta_{3}Y \right)
\; .
\end{eqnarray}
We observe, based on these results, that the $Q_{Z}$ generator of the GSW model is obtained whenever $\theta_{1}=0$. Thereby, we admit this solution for $\theta_{1}$ to recover all the interactions of the $Z$-boson with the fermions of the SM. The interaction of neutrinos and leptons with the $W^{\pm}$ are
recovered exactly like in the GSW model.
%
%
The $Q_{Z}$-generator emerges as in the usual Electroweak Model, but we have here another charge, $Q_{Z'}$, of the interaction between the fermions
and the $Z'$-boson. The values of $Q_{em}$, $I^{3}$, $Y=Q+Q'$ and the primitive charges $Q$ and $Q'$ are summarized in table (\ref{Table1}).
These values are due to a possible solution for the Abelian (chiral) anomaly to cancel out. Since the model is based on two Abelian subgroups,
there are six triangle graphs of the $Q$- and $Q'$-symmetries that contribute for the anomaly : $Q$ , $Q'$ , $Q^2 \, Q'$ , $Q \, Q'^2$ , $Q^3$ , $Q'^3$.
Therefore, the sum of all the charges $Q$ and $Q'$ in the table (\ref{Table1}) ensure the cancellation of the six triangle graphs. The necessary condition for an anomaly-free model is that the $\zeta_{i}$-fermions have no electric charge, {\it i. e.}, $Y=0$, with $Q=-Q'=+1/2$ for Left-component and $Q=-Q'=-1/2$ for Right-component. Furthermore, the neutrino right-component must also be added to guarantee the model to be free from anomalies.
Therefore, the Yukawa interactions introduced in the Higgs sector are gauge invariant with the $Q$- and $Q'$-charges in the table. As a consequence, $\zeta_{i}$-fermions do not interact with $Z$-boson and photon of the SM, but it does interact with the $Z'$-boson. All the interactions of the Electroweak Standard Model are reproduced in the model. The right-neutrino components and $\zeta_{i}$-fermions do not interact with the EW $Z$-boson, as it can be verified by the consulting the respective charges.
\begin{table}
\centering
\begin{tabular}{|l|l|l|l|l|l|}
\hline
\mbox{Fields} \& \mbox{particles} & $Q_{em}$ & $I^{3}$ & $Y$ & $Q$ & $Q'$ \\
\hline
\mbox{lepton-left} & $-1$ & $-1/2$ & $-1/2$ & $ 0 $ & $-1/2 $  \\
\hline
\mbox{neutrino-left} & $0$ & $+1/2$ & $-1/2$ & $ 0 $ & $ -1/2 $ \\
\hline
\mbox{lepton-right} & $-1$ & $0$ & $-1$ & $-1/2$ & $ -1/2 $ \\
\hline
\mbox{neutrino-right} & $0$ & $0$ & $0$ & $+1/2$ & $-1/2$ \\
\hline
$\zeta_{i}$-\mbox{fermions left} & $0$ & $0$ & $0$ & $-1/2$ & $+1/2$ \\
\hline
$\zeta_{i}$-\mbox{fermions right} & $0$ & $0$ & $0$ & $+1/2$ & $-1/2$ \\
\hline
\mbox{u-quark-left} & $+2/3$ & $+1/2$ & $+1/6$ & $0$ & $+1/6$  \\
\hline
\mbox{d-quark-left} & $-1/3$ & $-1/2$ & $+1/6$ & $0$ & $+1/6$ \\
\hline
\mbox{u-quark-right} & $+2/3$ & $0$ & $+2/3$ & $+1/2$ & $+1/6$ \\
\hline
\mbox{d-quark-right} & $-1/3$ & $0$ & $-1/3$ & $-1/2$ & $+1/6$ \\
\hline
$W^{\pm}$-\mbox{bosons} & $\pm \, 1$ & $\pm \, 1$ & $0$ & $0$ & $0$ \\
\hline
\mbox{neutral bosons} & $0$ & $0$ & $0$ & $0$ & $0$  \\
\hline
$\Xi$-\mbox{Higgs} & $0$ & $0$ & $0$ & $-1$ & $+1$ \\
\hline
$\Phi$-\mbox{Higgs} & $0$ & $-1/2$ & $+1/2$ & $+1/2$ & $0$ \\
\hline
\end{tabular}
\caption{The particle content for the $Z'$-model candidate at the TeV-scale physics.
The $Q$- and $Q'$-charges are such that anomalies cancel out.}\label{Table1}
\end{table}
%
%
The new interactions of the leptons, neutrinos and the $\zeta_{i}$-fermions with the $Z'$-boson are displayed in what follows :
\begin{equation}\label{LintZ'}
{\cal L}^{int}_{Z'}=- \, g_{Z'} \,
\bar{f} \, \, \slash{\!\!\!\!Z}' \left(g^{f}_{V}-g^{f}_{A}\gamma_{5}\right)f \; ,
\end{equation}
where $g_{Z'}:=e\csc(2\theta_{3})\sec\theta_{W}$, and we rewrite the $Z'$-interaction with the Left- and Right-components of fermions
$\Psi$ from (\ref{LintAZC}). Here, the $f$-fermions means the fermion fields with no quiral components. Furthermore, we define the coefficients
$g^{f}_{V}$ and $g^{f}_{A}$ as
\begin{eqnarray}
g^{f}_{V} &=& -Q^{f_{L}}-Q^{f_{R}}+\left(Y^{f_{L}}+Y^{f_{R}}\right)\sin^2\theta_{3}
\nonumber \\
g^{f}_{A} &=& -Q^{f_{L}}+Q^{f_{R}}+\left(Y^{f_{L}}-Y^{f_{R}}\right)\sin^2\theta_{3} \; .
\end{eqnarray}
The $f$-sum in (\ref{LintZ'}) runs to all $f$-fermions (no quiral components) of the model.
Thereby, we list all values of $g^{f}_{V}$ and $g^{f}_{A}$ following the charges in the table (\ref{Table1}) :
\begin{eqnarray}
g^{\ell_{i}}_{V} &=& -1+\frac{3}{2} \cos^2\theta_{3}
\hspace{0.3cm} , \hspace{0.3cm}
g^{\ell_{i}}_{A}= -\frac{1}{2} \cos^2\theta_{3}
\hspace{0.3cm} , \hspace{0.3cm}
\nonumber \\
g^{\nu_{i}}_{V} &=& -1+ \frac{1}{2} \cos^2\theta_{3}
\hspace{0.3cm} , \hspace{0.3cm}
g^{\nu_{i}}_{A}= \frac{1}{2} \cos^2\theta_{3}
\hspace{0.3cm} , \hspace{0.3cm}
\nonumber \\
g^{u}_{V} &=& -\frac{1}{2}+\frac{5}{6} \sin^2\theta_{3}
\hspace{0.3cm} , \hspace{0.3cm}
g^{u}_{A}= \frac{1}{2} \cos^2\theta_{3}
\hspace{0.3cm} , \hspace{0.3cm}
\nonumber \\
g^{d}_{V} &=& \frac{1}{2}-\frac{1}{6} \sin^2\theta_{3}
\hspace{0.3cm} , \hspace{0.3cm}
g^{d}_{A}= -\frac{1}{2} \cos^2\theta_{3}
\hspace{0.3cm} , \hspace{0.3cm}
\nonumber \\
g^{\zeta_{i}}_{V} &=& 0
\hspace{0.3cm} , \hspace{0.3cm}
g^{\zeta_{i}}_{A}= +1  \; .
\end{eqnarray}

We have then obtained the masses and physical fields of the $W^{\pm}$-, $Z$- and $Z'$-bosons,
and the interaction of $Z'$ with the fermion fields of the SM. The diagonalization of mixed sector of
RHNs with the set of $\zeta_{i}$-fermions shall be the issue of the next Section.

\section{The mixing between right-neutrinos and the $\zeta_{i}$-fermions}
\renewcommand{\theequation}{3.\arabic{equation}}
\setcounter{equation}{0}
After the SSB takes place, the neutrinos and the $\zeta_{i}$-fermions acquire
mass terms as displayed below:
\begin{eqnarray}\label{LpsichiM}
{\cal L}_{\nu_{i}-\zeta_{i}}^{\, 0} \!\!&=&\!\! \bar{\nu}_{iL} \, i \, \slash{\!\!\!\partial} \; \nu_{iL}
+\bar{\nu}_{iR} \, i \, \slash{\!\!\!\partial} \; \nu_{iR}
+\bar{\zeta}_{iL}\, i \, \slash{\!\!\!\partial} \; \zeta_{iL}
+\bar{\zeta}_{iR}\, i \, \slash{\!\!\!\partial} \; \zeta_{iR}
\nonumber \\
&&
\hspace{-0.4cm}
- \frac{v}{\sqrt{2}}
\, X_{ij} \, \bar{\nu}_{iL} \, \nu_{jR}
- \frac{v}{\sqrt{2}}
\, Y_{ij} \, \bar{\nu}_{iL} \, \zeta_{jR}
\nonumber \\
&&
\hspace{-0.4cm}
-\frac{u}{\sqrt{2}}
\, Z_{ij} \, \bar{\zeta}_{iL} \, \nu_{jR}
- \frac{u}{\sqrt{2}}
\, W_{ij} \, \bar{\zeta}_{iL} \, \zeta_{jR}
+\mbox{h. c.} \; .
\end{eqnarray}
This can be cast in the matrix form
\begin{eqnarray}\label{LnuzetaMatrix}
{\cal L}_{\nu_{i}-\zeta_{i}}^{\, 0}=
\bar{\chi}_{L} \, i \, \slash{\!\!\!\partial} \, \chi_{L}
+\bar{\chi}_{R} \, i \, \slash{\!\!\!\partial} \, \chi_{R}
-\bar{\chi}_{L} \, M \, \chi_{R} +\mbox{h. c.} \; ,
\hspace{0.2cm}
\end{eqnarray}
where we have defined the column-matrix $\chi_{L(R)}^{t}:=\left( \, \nu_{iL(R)} \; \; \zeta_{iL(R)} \, \right)$ and the mass matrix
is given by
\begin{eqnarray}\label{MassMatrixM}
M=\frac{1}{\sqrt{2}}
\left(
\begin{array}{cc}
v \, X & v \, Y
\\
\\
u \, Z & u \, W
\\
\end{array}
\right) \; .
\end{eqnarray}
Here, $X$, $Y$, $Z$ and $W$ are $3 \times 3$-matrices of elements $\{ \, X_{ij} \, , \, Y_{ij} \, , \, Z_{ij} \, , \, W_{ij} \, \}$, and thus,
the mass matrix is actually $6 \times 6$. We introduce the unitary transformation
\begin{eqnarray}\label{chiLRtransf}
\chi_{L(R)} \, \longmapsto \, \chi'_{L(R)} \!&=&\! S \, \chi_{L(R)} \; ,
\end{eqnarray}
where $S$ is the unitary matrix, $S^{\dagger} S={\uma}$. The neutrino-$\zeta$-fermion sector (\ref{LnuzetaMatrix})
is diagonal and the mass matrix, after the transformation (\ref{chiLRtransf}), satisfies the relation
\begin{eqnarray}\label{MD}
M_{D}=S \, M \, S^{\dagger}=
\left(
\begin{array}{cc}
M^{(\nu)} & 0
\\
0 & M^{(\zeta)} \\
\end{array}
\right)
\; .
\end{eqnarray}
To obtain the form of $S$, we begin with the most general form of a $U(2)$-matrix
\begin{eqnarray}
S=\left(
\begin{array}{cc}
e^{i \, \alpha} \, \cos\theta & e^{i \, \beta} \, \sin\theta \\
-e^{i \, (\gamma-\beta)} \, \sin\theta & e^{i \, (\gamma-\alpha)} \, \cos\theta \\
\end{array}
\right) \; ,
\end{eqnarray}
with four independent parameters : $\theta$ is mixing angle between the families of Right-Neutrino and $\zeta$-fermion, and three phases $\{ \, \alpha \, , \, \beta \, , \, \gamma \, \}$. The $M_{D}$-matrix is diagonal if the angles $\left\{ \, \theta \, , \, \alpha \, , \, \beta \, \right\}$ satisfy the relation
\begin{equation}\label{tantheta}
\left( v \, X - u \, W \right)\tan(2\theta)= 2 \, v \, Y \, e^{i \, (\alpha-\beta)}=2 \, u \, Z \, e^{-i \, \left(\alpha-\beta\right)} \; ,
\end{equation}
and the two $3 \times 3$ matrices of (\ref{MD}) are given by the eigenvalues
\begin{eqnarray}\label{MnuMzeta}
M^{(\nu)} \!\!&=&\!\! \frac{ v \, X + u \, W - \sqrt{\left(v \, X - u \, W \right)^{2}+4 v \, u \, Z \, Y }}{2\sqrt{2}} \; ,
\nonumber \\
M^{(\zeta)} \!\!&=&\!\! \frac{ v \, X + u \, W + \sqrt{\left(v \, X - u \, W \right)^{2}+4 v \, u \, Z \, Y }}{2\sqrt{2}} \; .
\hspace{0.6cm}
\end{eqnarray}
%
%
%
The diagonalization does not depend on the $\gamma$-phase, so that we can eliminate it.
%
%
It can be readily seen that, whenever $Y=Z=0$, we obtain the mass matrices $M^{(\nu)}=vX/\sqrt{2}$ and $M^{(\zeta)}=uW/\sqrt{2}$, respectively.
Here, we are in the $Z'$-scenario where the $u$- VEV scale satisfies
the condition $u \gg v=246 \, \mbox{GeV}$; so, we hope that the set of fermions
$\zeta_{i}=\left\{ \, \zeta_{1} \, , \, \zeta_{2} \, , \, \zeta_{3} \, \right\}$
should describe three particles heavier than any neutrino of the SM,
{\it i. e.}, it is reasonable to consider that $u \, W \gg v \, X$. Furthermore,
the elements $\{ \, Y_{ij} \, , \, Z_{ij} \, \}$ can be considered weaker coupling constants
due to the mixing of right-neutrinos with the sector of $\zeta_{i}$-fermions. Under these conditions,
the mass matrices (\ref{MnuMzeta}) can be written as corrections of
%
\begin{eqnarray}\label{autoveloresmfermions}
M^{(\nu)} \!&\simeq&\! \frac{v \, X}{\sqrt{2}} \left( \, {\uma} - \frac{ZY}{XW} \, \right) \; ,
\nonumber \\
M^{(\zeta)} \!&\simeq&\! \frac{u \, W}{\sqrt{2}} \left( \, {\uma} +  \frac{v}{u} \, \frac{ZY}{W^2} \, \right)
\simeq \frac{u \, W}{\sqrt{2}} \; .
\end{eqnarray}
%
%

%
The neutrino-$\zeta_{i}$-fermion sector is now free from mixed terms $\tilde{\nu}_{i}-\tilde{\zeta}_{i}$ in the basis $\tilde{\chi}_{L(R)}$ :
\begin{eqnarray}\label{Lnuzetalinha}
{\cal L}_{\tilde{\nu}_{i}-\tilde{\zeta}_{i}}^{\, 0} \!\!\!&=&\!\! \bar{\tilde{\nu}}_{iL} \, i \, \slash{\!\!\!\partial} \, \tilde{\nu}_{iL}
+\bar{\tilde{\nu}}_{iR} \, i \, \slash{\!\!\!\partial} \, \tilde{\nu}_{iR}
+\bar{\tilde{\zeta}}_{iL} \, i \, \slash{\!\!\!\partial} \, \tilde{\zeta}_{iL}
+\bar{\tilde{\zeta}}_{iR} \, i \, \slash{\!\!\!\partial} \, \tilde{\zeta}_{iR}
\nonumber \\
&&
\hspace{-0.5cm}
-\bar{\tilde{\nu}}_{iL} \, M_{ij}^{\, (\nu)} \, \tilde{\nu}_{jR}-\bar{\tilde{\zeta}}_{iL} \, M_{ij}^{\, (\zeta)} \, \tilde{\zeta}_{jR}
+\mbox{h. c.} \; .
\end{eqnarray}
Therefore, the mass matrices in (\ref{Lnuzetalinha}) can be independently diagonalized. The basis $\chi_{L(R)}$ can be
written in terms of $\tilde{\chi}_{L(R)}$ by the inverse transformation of (\ref{chiLRtransf}), {\it i. e.}, the transformation
$\left( \, \nu_{iL(R)} \, , \, \zeta_{iL(R)} \, \right) \mapsto \left(\, \tilde{\nu}_{iL(R)} \, , \, \tilde{\zeta}_{iL(R)} \, \right)$
\begin{eqnarray}
\nu_{iL(R)} \!\!&=&\!\! e^{-i \, \alpha} \, \cos\theta \, \tilde{\nu}_{iL(R)} - e^{-i \, \beta} \, \sin\theta \, \tilde{\zeta}_{iL(R)} \; ,
\nonumber \\
\zeta_{iL(R)} \!\!&=&\!\! e^{-i \, \beta} \, \sin\theta \, \tilde{\nu}_{iL(R)} + e^{i \, \alpha} \, \cos\theta \, \tilde{\zeta}_{iL(R)} \; .
\end{eqnarray}
The interactions $\nu-\ell-W^{\pm}$ are written in the new basis $\left(\, \tilde{\nu}_{L(R)} \, , \, \tilde{\zeta}_{L(R)} \, \right)$, such that the neutrino-lepton-$W^{\pm}$ interaction is given by
\begin{equation}\label{LintnnuWell}
{\cal L}_{\tilde{\nu}_{i}-\ell_{i}-W}^{int}=-\frac{g \cos\theta}{\sqrt{2}} \, \, \bar{\tilde{\nu}}_{iL} \, \, \slash{\!\!\!\!W}^{+} \, \ell_{iL}
+\mbox{h. c.}
\end{equation}
where the $\alpha$-phase has been absorbed into the left-neutrino field, and $\cos\theta\simeq 1$ for $\theta \ll 1$.
Therefore, this change of basis does not affect the already-known neutrino-lepton-$W^{\pm}$ interaction of the GSW model.
The mass basis of the Dirac neutrinos is introduced via a unitary transformation of the $\nu_{L(R)}$-fields,
then the Pontecorvo-Maki-Nakagawa-Sakata (PMNS) matrix emerges in the interaction (\ref{LintnnuWell}) with
only one Dirac phase for Dirac Right-Neutrinos. Furthermore, there are three independent angles in the PMNS matrix to mix the neutrino fields, as usually.
If we introduce the unitary transformation $\tilde{\nu}_{L(R)} \, \longmapsto \, \nu'_{L(R)}=S_{L(R)} \, \tilde{\nu}_{L(R)}$,
the neutrino mass matrix in (\ref{Lnuzetalinha}) is diagonal in the mass basis $\nu'_{L(R)}$, in which we have $\bar{\tilde{\nu}}_{L} \, M^{(\nu)} \, \tilde{\nu}_{R}=\bar{\nu}'_{L} \, M_{D}^{(\nu)} \, \nu'_{R}$, where $M_{D}^{(\nu)}$ is given by
\begin{equation}\label{MDneutrinos}
M_{D}^{(\nu)}=S_{L}\frac{v \, X}{\sqrt{2}} \left( \, {\uma} - \frac{ZY}{XW} \, \right)S_{R}^{\dagger}
=\left(
\begin{array}{ccc}
M_{\nu'_{e}} & 0 & 0 \\
0 & M_{\nu'_{\mu}} & 0 \\
0 & 0 & M_{\nu'_{\tau}} \\
\end{array} \right) \; .
\end{equation}
Since the neutrino masses are measured through their oscillations, the transition probabilities depend on the subtraction of the
squared masses. In the case of the electron- and muon-neutrinos, this subtraction is $\Delta M_{\nu'_{e}-\nu'_{\mu}}^{2}:=|M_{\nu'_{e}}^2-M_{\nu'_{\mu}}^2|\simeq \left( \, 7.53 \, \pm \, 0.18 \, \right) \times 10^{-5} \, \mbox{eV}^{2}$ \cite{ArakiPRL2005}. Thus, the subtraction of squared coupling constants are extremely weak,
$\Delta X_{\nu'_{e}-\nu'_{\mu}}^{2}:=||X_{\nu'_{e}}|^2-|X_{\nu'_{\mu}}|^2|\simeq 2.5 \times 10^{-27}$.
The mixing between $\tau$- and muon- neutrino yield the squared subtraction
$\Delta M_{\nu'_{\tau}-\nu'_{\mu}}^{2}:=|M_{\nu'_{\tau}}^2-M_{\nu'_{\mu}}^2| \simeq \left( \, 2.44 \, \pm \, 0.06 \, \right) \times 10^{-3} \, \mbox{eV}^{2}$, then we estimate $\Delta X_{\nu'_{\tau}-\nu'_{\mu}}^{2}:=||X_{\nu'_{\tau}}|^2-|X_{\nu'_{\mu}}|^2|\simeq 8 \times 10^{-26}$.
This estimation helps us to obtain the corresponding values of mixed Yukawa coupling constants $Y$ and $Z$,
but we need to define a range for the $\zeta_{i}$-masses.
The $\zeta_{i}$-fermions interact weakly with the leptons according to
\begin{equation}\label{Lintzetatilell}
{\cal L}_{\tilde{\zeta}-\ell-W}^{int}=\frac{g \sin\theta}{\sqrt{2}} \, \, \bar{\tilde{\zeta}}_{iL} \, \, \slash{\!\!\!\!W}^{+} \, \ell_{iL}
+\mbox{h. c.}
\end{equation}
where the $\beta$-phase has been absorbed into the $\tilde{\zeta}_{iL}$-fields. The leptonic sector is diagonalized like
in the SM. The mass term is $-v \, \bar{\ell}_{L} \, G^{(\ell)} \, \ell_{R}/\sqrt{2}$, and it can be diagonalized by means of the
unitary transformations
%
$\ell_{L(R)} \, \longmapsto \, \ell'_{L(R)}=U_{L(R)} \, \ell_{L(R)}$,
%
where $U_{L}^{\dagger} \, U_{L}=U_{R}^{\dagger} \, U_{R}={\uma}$, the mass matrix diagonal is
$M_{D}^{(\ell)}=U_{L} \, v \, G^{(\ell)}/\sqrt{2} \, U_{R}^{\dagger}=\mbox{diag}\left( \, M_{e} \, , \, M_{\mu} \, , \, M_{\tau} \, \right)$.
Analogously, the diagonalization of $\zeta_{i}$- mass matrix is carried out by another unitary transformation,
that we denote by $\tilde{\zeta}_{L(R)} \, \longmapsto \, \zeta'_{L(R)}=V_{L(R)} \, \tilde{\zeta}_{L(R)}$, where $V_{L}^{\dagger} \, V_{L}=V_{R}^{\dagger} \, V_{R}={\uma}$, and we obtain the diagonal mass matrix below:
\begin{eqnarray}
M_{D}^{\, (\zeta)}=V_{L} \, \frac{u \, W}{\sqrt{2}} \, V_{R}^{\dagger}
= \left(
\begin{array}{ccc}
M_{\zeta'_{1}} & 0 & 0 \\
0 & M_{\zeta'_{2}} & 0 \\
0 & 0 & M_{\zeta'_{3}} \\
\end{array} \right) \; .
\end{eqnarray}
So, we can write the interactions (\ref{Lintzetatilell})
in the mass basis $\left\{ \, \zeta'_{L} \, , \, \ell'_{L} \, \right\}$ as follows:
\begin{equation}\label{Intzeta'ell'W}
{\cal L}_{\zeta'_{i}-\ell'_{i}-W}^{int}=\frac{g \sin\theta}{\sqrt{2}} \, \, \bar{\zeta}'_{iL} \, \, \slash{\!\!\!\!W}^{+} \, V_{ij} \, \ell'_{jL}
+\mbox{h. c.}
\end{equation}
where the most general $V$-unitary matrix $V:=V_{L} \, U_{L}^{\dagger}$ is parameterized by
\begin{widetext}
\begin{equation}\label{PMNSMatrix}
V=\left(
\begin{array}{ccc}
c_{12}c_{13} & s_{12}c_{13} & s_{13} \, e^{-i\delta} \\
-s_{12}c_{23}-c_{12}s_{23}c_{13}e^{i\delta} & c_{12}c_{23}-s_{12}s_{23}s_{13}e^{i\delta} & s_{23}c_{13} \\
s_{12}s_{23}-c_{12}c_{23}s_{13}e^{i\delta} & -c_{12}s_{23}-s_{12}c_{23}s_{13}e^{i\delta} & c_{23}c_{13} \\
\end{array}
\right) .
\end{equation}
\end{widetext}
It has the same structure as the Cabibbo-Kobayashi-Maskawa (CKM) matrix : it displays three mixing angles
$\left\{ \, \theta_{12} \, , \, \theta_{13} \, , \, \theta_{23} \, \right\}$ and only one Dirac $\delta$-phase,
since we do not introduce $\zeta_{i}$-Majorana fermions. In (\ref{PMNSMatrix}), we simply the
sines and cosines of the angles as : $\cos\theta_{ij}=c_{ij}$ and $\sin\theta_{ij}=s_{ij}$.
The $\zeta'_{i}$-masses depend on the $u$-VEV scale, so it must exhibit a heavier fermion content
in comparison with the SM fermions. The recent simulations of CMS-Collaboration point out to
DM fermion content with mass of order $0.55 \, \mbox{TeV}$,
when the mediator $Z'$ is the axial-vector with a mass around the $2 \, \mbox{TeV}$ \cite{CMS20172}.
Therefore, we take here $M_{\zeta'_{1}}=0.5 \, \mbox{TeV}$, and if we use $u=2.8 \, \mbox{TeV}$,
the estimation for $W_{\zeta'_{1}}$-Yukawa constant is $|W_{\zeta'_{1}}| \simeq 0.28$.
The two others heavy fermions $\zeta'_{2}$ and $\zeta'_{3}$ can be DM particles in the mass range of
$>0.5 \, \mbox{TeV}$, so we choose $M_{\zeta'_{2}}=0.8 \, \mbox{TeV}$ and $M_{\zeta'_{3}}=1 \, \mbox{TeV}$,
so the correspondent coupling constants has the values $|W_{\zeta'_{2}}| \simeq 0.4$ and $|W_{\zeta'_{3}}| \simeq 0.5$, respectively.
Using the uncertainty on $\Delta M_{\nu'_{e}-\nu'_{\mu}}^{2}$, the correction to the neutrinos masses gives the upper bound
\begin{eqnarray}
\frac{\Delta Y_{\nu'_{e}-\nu'_{\mu}}\Delta Z_{\nu'_{e}-\nu'_{\mu}}}{\Delta X_{\nu'_{e}-\nu'_{\mu}}W_{\zeta'_{1}}} \lesssim 0.15 \; ,
\end{eqnarray}
and $\Delta Y \simeq \Delta Z$, we obtain $\Delta Y_{\nu'_{e}-\nu'_{\mu}}\simeq\Delta Z_{\nu'_{e}-\nu'_{\mu}}\simeq 8.3 \, \times \, 10^{-8}$.
Under these conditions and the previous bounds, the $\theta$-mixing angle in (\ref{tantheta}) turns out to be extremely small :
$\tan\theta\simeq-9\, \times \, 10^{-8}$.

The interaction of the $\zeta_{i}$-fermions with the $Z'$-boson is not affected by the change of basis dictated by the masses.
Another important feature is that the interaction (\ref{Intzeta'ell'W}) connects the fermion sector of the SM
with a set of fermions candidate to dark sector via $W^{\pm}$-bosons. Since $\theta \ll 1$, the $\theta$-angle
rules the magnitude of (\ref{Intzeta'ell'W}), {\it i. e.}, $\sin\theta \simeq \theta \simeq-9\, \times \, 10^{-8}$.
This vertex is represented by the diagram below.
\begin{figure}[!h]
\begin{center}
\newpsobject{showgrid}{psgrid}{subgriddiv=1,griddots=10,gridlabels=6pt}
\begin{pspicture}(5,1)(11,2.5)
\psset{arrowsize=0.2 2}
\psset{unit=0.8}
%
%
\pscoil[coilarm=0,coilaspect=0,coilwidth=0.2,coilheight=1.0,linecolor=black](6.5,1.05)(6.5,3)
\psline[linecolor=black,linewidth=0.5mm]{-}(5,1)(8,1)
\psline[linecolor=black,linewidth=0.5mm]{->}(5,1)(6,1)
\psline[linecolor=black,linewidth=0.5mm]{->}(7,1)(7.55,1)
\put(6.8,2.8){\large$W^{\pm}$}
\put(4.95,1.3){\large$\bar{\zeta}'_{i}$}
\put(7.8,1.3){\large$\ell'_{j}$}
\put(8.5,1.1){\large $\Gamma_{ij}^{\, \mu}=- \, \frac{i \, g \,\theta \, V_{ij}}{2\sqrt{2}} \gamma^{\mu}\left(1- \gamma_{5}\right) .$}
\end{pspicture}
%
%
%
\end{center}
\end{figure}
\noindent
Therefore, this vertex yields an important contribution to $\zeta_{i}$-fermions magnetic dipole momentum at the one-loop approximation.
The weakly coupling constant that emerges here is $g \, \theta \simeq-6 \times 10^{-9}$. The interaction (\ref{Intzeta'ell'W})
also violates the CP-symmetry due to the $\delta$-phase in the mixing matrix (\ref{PMNSMatrix}). This extra violation disappears
when the $\theta$-mixing coupling goes to zero.

%
%

%
\section{On the $Z'$-phenomenology}
\renewcommand{\theequation}{4.\arabic{equation}}
\setcounter{equation}{0}

\subsection{The $Z'$-decay into fermions : $Z' \, \rightarrow \, \bar{f} \, f$}
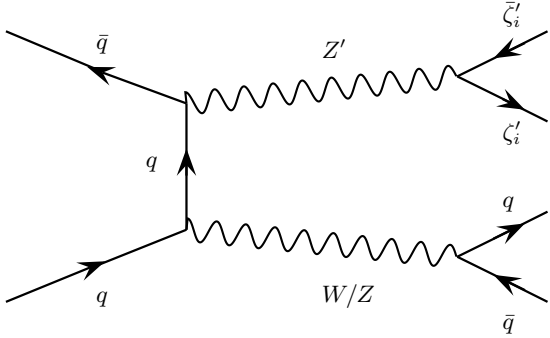
\begin{figure}[!h]
\begin{center}
\newpsobject{showgrid}{psgrid}{subgriddiv=1,griddots=10,gridlabels=6pt}
\begin{pspicture}(0,-1)(2.5,3.5)
\psset{arrowsize=0.2 2}
\psset{unit=1.2}
%
%
\pscoil[coilaspect=0,coilarm=0,coilwidth=0.25,coilheight=1.3,linecolor=black](0,0.3)(2.99,0)
\pscoil[coilaspect=0,coilarm=0,coilwidth=0.25,coilheight=1.3,linecolor=black](0,1.7)(2.99,2)
\put(1.5,2.2){$Z'$}
\put(1.5,-0.5){$W/Z$}
%
%
\psline[linecolor=black,linewidth=0.3mm]{->}(0,0.3)(0,1.2)
\psline[linecolor=black,linewidth=0.3mm]{-}(0,0.9)(0,1.7)
\put(-0.45,1){$q$}
%
%
\psline[linecolor=black,linewidth=0.3mm]{->}(-2,-0.5)(-0.9,-0.05)
\psline[linecolor=black,linewidth=0.3mm]{-}(-1,-0.1)(0,0.3)
\put(-1,2.3){$\bar{q}$}
\psline[linecolor=black,linewidth=0.3mm]{->}(0,1.7)(-1.1,2.105)
\psline[linecolor=black,linewidth=0.3mm]{-}(-1,2.08)(-2,2.5)
\put(-1,-0.5){$q$}
%
%
\psline[linecolor=black,linewidth=0.3mm](3,2)(3.75,2.37)
\psline[linecolor=black,linewidth=0.3mm]{<-}(3.4,2.2)(4,2.5)
\put(3.5,2.6){$\bar{\zeta}'_{i}$}
\psline[linecolor=black,linewidth=0.3mm]{->}(3,2)(3.75,1.63)
\psline[linecolor=black,linewidth=0.3mm](3.5,1.75)(4,1.5)
\put(3.5,1.3){$\zeta'_{i}$}
%
%
\psline[linecolor=black,linewidth=0.3mm]{->}(3,0)(3.75,0.37)
\psline[linecolor=black,linewidth=0.3mm](3.4,0.2)(4,0.5)
\put(3.5,0.55){$q$}
\psline[linecolor=black,linewidth=0.3mm](3,0)(3.75,-0.37)
\psline[linecolor=black,linewidth=0.3mm]{<-}(3.4,-0.2)(4,-0.5)
\put(3.5,-0.8){$\bar{q}$}
%
%
%
%
%
%
%
%
%
%
%
\end{pspicture}
%
%
\caption{\scshape{The $Z'$-decay into any pair $\bar{\zeta}'_{i}-\zeta'_{i}$ of the set of $\zeta'_{i}$-fermions.
The cascade effect as a possible DM detection via $W$- or $Z$-monojets.}} \label{Z'Decay}
\end{center}
\end{figure}

The search for DM has been motivated by study of energetic jets or a hadronically decay into the W- or Z-boson at $\sqrt{s}=13 \, \mbox{TeV}$ \cite{CMS20172}. The recent $Z'$-phenomenology points to the cascade effects at the tree-level using the CMS data for the pp-collision at $\sqrt{s}=13 \, \mbox{TeV}$. Motivated by these tree-level effects, we shall obtain a general expression for the $Z'$ decay width into any $f$-fermion of the model.
Then, using the previous rules and quantum field-theoretic results,
the decay width of $Z'$ into any $f$-fermion is given by
\begin{eqnarray}
\Gamma(Z' \rightarrow \bar{f} \, f ) \!&=&\! \frac{g_{Z'}^{2} M_{Z'} }{24\pi}
\left( |g^{f}_{V}|^2+|g^{f}_{A}|^2 \right)
\times
\nonumber \\
&&
\hspace{-0.5cm}
\times \,
\sqrt{1-\frac{4M_{f}^{\, 2}}{M_{Z'}^{\, 2}}} \left( 1
- \frac{3M_{f}^{\, 2}}{4M_{Z'}^{\, 2}} \right) \, ,
\end{eqnarray}
where $M_{Z'} > 2 M_{f}$, for $f=\left\{ \, \ell_{i} \, , \, \nu_{i} \, , \, \zeta_{i} \, \right\}$ or quarks.
The decay widths into a lepton pair and a neutrino pair provide the right-components contribution and are given by
\begin{eqnarray}\label{Z'Decayll}
\Gamma(Z' \rightarrow \bar{\ell}_{i} \, \ell_{i} ) \!&=&\! \frac{g_{Z'}^2 M_{Z'} }{48\pi}
\left(1-4\sin^2\theta_{3}+5 \sin^4\theta_{3} \right) \; ,
\nonumber \\
\Gamma(Z' \rightarrow \bar{\nu}_{i} \, \nu_{i} ) \!&=&\! \frac{g_{Z'}^2 M_{Z'} }{48\pi}
\left( 1+ \sin^4\theta_{3} \right) \; ,
\end{eqnarray}
%
%
%
and, if we use that $\theta_{3}=45^{0}$, we obtain $\Gamma(Z' \rightarrow \bar{\ell}_{i} \, \ell_{i} )=0.38 \, \mbox{GeV}$ and
$\Gamma(Z' \rightarrow \bar{\nu}_{i} \, \nu_{i} )= 1.9 \, \mbox{GeV}$. Notice that we have used that $M_{Z'}\gg \left\{ \, 2 \, M_{\ell_{i}} \, , \, 2 \, M_{\nu_{i}} \, \right\}$ for leptons and neutrinos. The processes of the $Z'$-decay can be useful to search for DM through the mono-V jets channels associated with the electroweak bosons $W$ or $Z$. The observation of these final states could be interpreted as a DM
particle content, that here we identify as the $\zeta'_{i}$-fermions. The diagram for this effect is illustrated in figure (\ref{Z'Decay}).
%
%
%

\noindent
Therefore, the result for the total decay width of $Z'$ into the $\zeta_{i}$-family is given by the sum
\begin{eqnarray}
\Gamma(Z' \rightarrow \bar{\zeta}' \, \zeta')=\sum_{i=1}^{3}\Gamma\left( \, Z' \rightarrow \bar{\zeta}'_{i} \, \zeta'_{i} \, \right) \; ,
\end{eqnarray}
in which for a particular $\zeta'_{i}$, it is shown to be given by
%
\begin{eqnarray}
\Gamma(Z' \rightarrow \bar{\zeta}'_{i} \, \zeta'_{i})=\frac{g_{Z'}^2 M_{Z'}}{24\pi}
\sqrt{1-\frac{4M_{\zeta'_{i}}^{\, 2}}{M_{Z'}^{\, 2}}} \left( 1
- \frac{3M_{\zeta'_{i}}^{\, 2}}{4M_{Z'}^{\, 2}} \right) \, .
\end{eqnarray}
Here, the condition $M_{Z'} > 2 \, M_{\zeta'_{i}}$ must be satisfied for any $\zeta'_{i}$-fermion.
Using the previous values $M_{Z'}= 2 \, \mbox{TeV}$ and $M_{\zeta'_{1}}=0.55 \, \mbox{TeV}$,
the $Z'$-width decay rate is
\begin{eqnarray}
\Gamma(Z' \rightarrow \bar{\zeta}'_{1} \, \zeta'_{1}) \simeq \frac{18 \, \mbox{GeV} }{\sin^2(2\theta_{3})} \; .
\end{eqnarray}
%
%
%
In the case of $\theta_{3}=45^{o}$, the decay width is
$\Gamma(Z' \rightarrow \bar{\zeta}'_{1} \, \zeta'_{1}) \simeq 18 \, \mbox{GeV}$, and the $Z'$-decay time in this process is estimated by
\footnote{We have used the conversion formula $1 \, \mbox{TeV}=1.52 \times 10^{27} \, \mbox{s}^{-1}$ in the natural units $\hbar=c=1$. }
\begin{eqnarray}
\tau(Z' \rightarrow \bar{\zeta}'_{1} \, \zeta'_{1})=\frac{1}{\Gamma(Z' \rightarrow \bar{\zeta}'_{1} \, \zeta'_{1})}\simeq 3.7 \times 10^{-26} \, \mbox{s} \; .
\end{eqnarray}

The possible $Z'$-decays into quarks, {\it i. e.}, $Z' \rightarrow \bar{q} \, q$, have also been the object of a phenomenological analysis at the CMS Collaboration,
see \cite{CMS20172}. $Z'$-decays into the first generation, {\it i. e.},  $Z' \, \rightarrow \, \bar{u} \, u$ and $Z' \, \rightarrow \, \bar{d} \, d$,
present the decay widths below :
\begin{eqnarray}
\Gamma(Z' \rightarrow \bar{u} \, u) \!\! &=& \!\! \frac{g_{Z'}^{2} M_{Z'}}{48\pi}
\left( 1-\frac{8}{3} \sin^2\theta_{3}+\frac{17}{9} \sin^4\theta_{3} \right)
\; ,
\nonumber \\
\Gamma(Z' \rightarrow \bar{d} \, d) \!\! &=& \!\! \frac{g_{Z'}^2 M_{Z'}}{48\pi}
\left(1-\frac{4}{3} \sin^2\theta_{3}+\frac{5}{9} \sin^4\theta_{3} \right) \, ,
\hspace{0.8cm}
\end{eqnarray}
where we have used that $M_{Z'} \gg m_{u}$ and $M_{Z'} \gg m_{d}$. Using the $\theta_{3}$-angle of $\theta_{3}=45^{o}$, we obtain the decay widths at the $\mbox{GeV}$-scale :
\begin{eqnarray}
\Gamma(Z' \rightarrow \bar{u} \, u) \simeq 2 \, \mbox{GeV}
\hspace{0.2cm} , \hspace{0.2cm}
\Gamma(Z' \rightarrow \bar{d} \, d) \simeq 7 \, \mbox{GeV} \; .
\end{eqnarray}
\subsection{The $F$-Higgs decays}
%

\begin{figure}[!h]
\begin{center}
\newpsobject{showgrid}{psgrid}{subgriddiv=1,griddots=10,gridlabels=6pt}
\begin{pspicture}(-3,-0.7)(6,3)
\psset{arrowsize=0.2 2}
\psset{unit=1.2}
%
%

%
\pscoil[coilaspect=0,coilarm=0,coilwidth=0.25,coilheight=1.3,linecolor=black](0,0.9)(1.5,1)
\put(0.8,1.3){$Z'$}
%
%
%
%
%
\psline[linecolor=black,linewidth=0.3mm](-0.65,0.3)(0,0.9)
\psline[linecolor=black,linewidth=0.3mm]{->}(-1.5,-0.5)(-0.5,0.45)
\put(-0.8,1.9){$\bar{q}$}
\psline[linecolor=black,linewidth=0.3mm]{->}(0,0.9)(-0.85,1.7)
\psline[linecolor=black,linewidth=0.3mm](-0.45,1.33)(-1.5,2.3)
\put(-0.8,-0.1){$q$}
%
%
%
\psline[linestyle=dashed,linewidth=0.5mm](1.5,1)(3,2)
\psline[linestyle=dashed,linewidth=0.5mm](1.5,1)(3,0)
\put(2.1,1.7){$F$}
\put(2.1,0){$F$}
%
%
\psline[linecolor=black,linewidth=0.3mm]{->}(3,2)(3.75,2.37)
\psline[linecolor=black,linewidth=0.3mm](3.4,2.2)(4,2.5)
\put(3.5,2.6){$\ell_{i}^{-}$}
\psline[linecolor=black,linewidth=0.3mm](3,2)(3.75,1.63)
\psline[linecolor=black,linewidth=0.3mm]{<-}(3.5,1.75)(4,1.5)
\put(3.5,1.2){$\ell_{i}^{+}$}
%
%
\psline[linecolor=black,linewidth=0.3mm]{->}(3,0)(3.75,0.37)
\psline[linecolor=black,linewidth=0.3mm](3.4,0.2)(4,0.5)
\put(3.5,0.55){$\ell_{i}^{-}$}
\psline[linecolor=black,linewidth=0.3mm](3,0)(3.75,-0.37)
\psline[linecolor=black,linewidth=0.3mm]{<-}(3.4,-0.2)(4,-0.5)
\put(3.5,-0.8){$\ell_{i}^{+}$}
%
%
%
%
%
%
%
%
%
%
%
\end{pspicture}
%
%
\caption{\scshape{The leading-order Feynman diagram for the cascade decay of the $Z'$-resonance into a four-lepton final state.}} \label{Z'4leptons}
\end{center}
\end{figure}
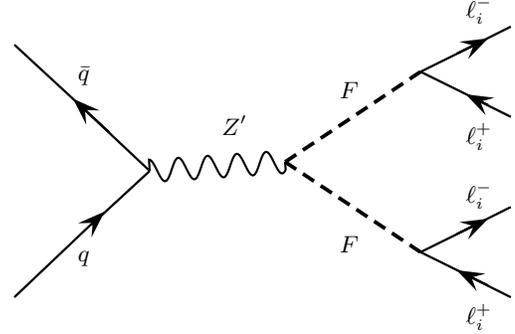

The $Z'$-decay into scalars has a considerable phenomenological interest for the
study of $Z'$-resonance to the final four-lepton state \cite{CMS20171}.
On the other hand, the $F$-Higgs decay into leptons pairs $Z' \, \rightarrow \, 4 \, \ell_{i}$.
The process is illustrated at the tree-level as shown in figure (\ref{Z'4leptons}).
%
%

\noindent

Moreover, the decay process $Z' \, \rightarrow \, F \, F$ cannot be described by the model due to action of couplings in the covariant derivative of the Higgs sector.
For example, after the SSB, the interaction of $F$-Scalar field with the $Z'$-boson is given by
\begin{eqnarray}
{\cal L}_{F-Z'}^{\, int}=\frac{M_{Z'}^{2}}{u} \, F \, Z_{\mu}^{\prime} Z^{\prime \mu}
+\frac{1}{2} \, \frac{M_{Z'}^{2}}{u^2} \, F^{2} \, Z_{\mu}^{\prime} Z^{\prime \mu} \; ,
\end{eqnarray}
where we have the possible vertex $F \, Z' \, Z'$ and $F \, F \, Z' \, Z'$. Using the usual rules of QFT, a process possible described by the $F-Z'$ sector is the decay $F \, \rightarrow \, Z' \, Z'$, that has the following decay width :
\begin{eqnarray}
\Gamma(F \, \rightarrow \, Z' \, Z') \!&=&\! \frac{g_{Z'}^{2} M_{Z'}}{16\pi} \frac{M_{Z'}}{M_{F}}
\left(1-\frac{4 M_{Z'}^{2}}{M_{F}^2} \right)^{1/2}
\times
\nonumber \\
&&
\hspace{-0.5cm}
\times
\left(3-\frac{M_{F}^{2}}{M_{Z'}^2}+\frac{M_{F}^{4}}{4M_{Z'}^4} \right) \, ,
\hspace{0.4cm}
\end{eqnarray}
where it is restricted by the condition $M_{F} > 2 \, M_{Z'}$.
However, other cascade effects can be described by the $Z'-F$-interaction. For example, the decay process in which the $Z'$ decays indirectly
into two final states of leptons and other two final states of $\zeta_{i}$-fermions,{\it i. e.}, $Z' \, \rightarrow \, 2 \, \ell_{i} + 2 \, \zeta'_{i}$.
This process is illustrated in figure (\ref{Z'2leptons2zeta}).
%
%
\begin{figure}[!h]
\begin{center}
\newpsobject{showgrid}{psgrid}{subgriddiv=1,griddots=10,gridlabels=6pt}
\begin{pspicture}(-3,-1)(6,3.3)
\psset{arrowsize=0.2 2}
\psset{unit=1.2}
%
%
%
\pscoil[coilaspect=0,coilarm=0,coilwidth=0.25,coilheight=1.3,linecolor=black](0,0.9)(1.5,1)
\put(0.8,1.3){$Z'$}
%
%
%
%
%
\psline[linecolor=black,linewidth=0.3mm](-0.65,0.3)(0,0.9)
\psline[linecolor=black,linewidth=0.3mm]{->}(-1.5,-0.5)(-0.5,0.45)
\put(-0.8,1.9){$\bar{q}$}
\psline[linecolor=black,linewidth=0.3mm]{->}(0,0.9)(-0.85,1.7)
\psline[linecolor=black,linewidth=0.3mm](-0.45,1.33)(-1.5,2.3)
\put(-0.8,-0.1){$q$}
%
%
%
\psline[linestyle=dashed,linewidth=0.5mm](1.5,1)(3,0)
\pscoil[coilaspect=0,coilarm=0,coilwidth=0.25,coilheight=1.3,linecolor=black](1.5,1)(3,2)
%
%
\put(2.1,1.8){$Z'$}
\put(2.1,0){$F$}
%
%
\psline[linecolor=black,linewidth=0.3mm]{->}(3,2)(3.75,2.37)
\psline[linecolor=black,linewidth=0.3mm](3.4,2.2)(4,2.5)
\put(3.5,2.6){$\ell_{i}^{-}$}
\psline[linecolor=black,linewidth=0.3mm](3,2)(3.75,1.63)
\psline[linecolor=black,linewidth=0.3mm]{<-}(3.5,1.75)(4,1.5)
\put(3.5,1.2){$\ell_{i}^{+}$}
%
%
\psline[linecolor=black,linewidth=0.3mm]{->}(3,0)(3.75,0.37)
\psline[linecolor=black,linewidth=0.3mm](3.4,0.2)(4,0.5)
\put(3.5,0.55){$\zeta_{i}^{'-}$}
\psline[linecolor=black,linewidth=0.3mm](3,0)(3.75,-0.37)
\psline[linecolor=black,linewidth=0.3mm]{<-}(3.4,-0.2)(4,-0.5)
\put(3.5,-0.8){$\zeta_{i}^{'+}$}
%
%
%
%
%
%
%
%
%
%
%
\end{pspicture}
%
%
\caption{\scshape{The leading-order Feynman diagram for the cascate decay of $Z'$ resonance to a four-lepton final state.}} \label{Z'2leptons2zeta}
\end{center}
\end{figure}
\noindent
In this case, we have part of the final state described by the decay width (\ref{Z'Decayll}).
The other process in the final state is the $F$-decay into the $\zeta'_{i}$-fields, which we denote as
$F \, \rightarrow \, \bar{\zeta}' \, \zeta'$.  The $F$-scalar field interacts with the $\zeta'_{i}$-fields
by means of the expression
\begin{eqnarray}
{\cal L}^{int}_{F-\bar{\zeta}' \, \zeta'}=- \sum_{i,j=1}^{3} \frac{|W_{\zeta'_{i}}|}{\sqrt{2}} \, F \, \bar{\zeta}'_{i} \, \zeta'_{j} \; .
\end{eqnarray}
Thus, the total decay width is given by
\begin{eqnarray}
\Gamma(F \, \rightarrow \, \bar{\zeta}' \, \zeta')=\sum_{i=1}^{3} \, \Gamma\left(F \, \rightarrow \, \bar{\zeta}'_{i} \, \zeta'_{i}\right) \; ,
\end{eqnarray}
where we obtain the decay width for each $\zeta_{i}$-fermion
\begin{eqnarray}
\Gamma(F \, \rightarrow \, \bar{\zeta}'_{i} \, \zeta'_{i} ) \!\!&=&\!\! M_{F} \, \frac{|W_{\zeta'_{i}}|^2}{8\pi}
\left[ \left(1- \frac{2M_{\zeta'_{i}}^2}{M_{F}^{2}}\right)
\, \times
\right.
\nonumber \\
&&
\hspace{-1cm}
\left.
\times
\sqrt{1-\frac{4M_{\zeta'_{i}}^2}{M_{F}^{2}}}
-\frac{2 \, M_{\zeta'_{i}}^{2}}{M_{F}^{2}}\left(1- \frac{M_{\zeta'_{i}}^2}{M_{F}^{2}} \right)
\right] \; ,
\hspace{0.6cm}
\end{eqnarray}
where $M_{F} > 2 \, M_{\zeta'_{i}} \, (i=1,2,3)$. If we use the $\zeta'_{1}$ mass of $M_{\zeta'_{1}}=0.5 \, \mbox{TeV}$, $|W_{\zeta'_{1}}|=0.28$ and the range (\ref{MassH1u8GEV}) for the F-mass, we obtain the estimative :
\begin{eqnarray}
0.5 \, \mbox{GeV} < \Gamma(F \, \rightarrow \, \bar{\zeta}'_{1} \, \zeta'_{1} ) < 10.3 \, \mbox{GeV} \; .
\end{eqnarray}
%

%
%

%
\section{On the MDM of the $\zeta_{i}$-fermions}
\renewcommand{\theequation}{5.\arabic{equation}}
\setcounter{equation}{0}

In this Section, we investigate the magnetic properties of $\zeta_{i}$-fermions through
their Magnetic Dipole Momentum (MDM)
The mixing of $\zeta_{i}$ with the right-neutrino components motivates us to understand if the $\zeta_{i}$-MDM
depends on its mass, as it happens in the case of neutrinos. For a review on the neutrinos' MDMs, go to the references \cite{BellPRL2005,BellPLB2006,Balantekin2006,Bhattacharya2004,StudenikinJP2016}. We start off with the field equations
in the mixed basis to obtain the so-called Transition Dipole Momenta (TDM). The Dirac equations for the
$\nu_{i}$-neutrinos and $\zeta_{i}$-fermions from (\ref{LpsichiM}) in momentum space are given by
\begin{eqnarray}\label{eqnuzeta}
\left( \, \slash{\!\!\!p} - m_{\nu_{i}} \, \right) u_{\nu_{i}}(p)-M_{ij}^{RL} \, u_{\zeta_{j}}(p) &=& 0 \; ,
\nonumber \\
\left( \, \slash{\!\!\!p} - m_{\zeta_{i}} \, \right) u_{\zeta_{i}}(p)-M_{ij}^{LR} \, u_{\nu_{j}}(p) &=& 0
\; ,
\end{eqnarray}
where $m_{\nu_{i}}$ and $m_{\zeta_{i}}$ are the masses if we consider the mixed coupling constants $Y_{ij}=Z_{ij} \, \rightarrow \, 0$.
The functions $u_{\nu_{i}}(p)$ and $u_{\zeta_{i}}(p)$ are the wave plane amplitudes of $\nu_{i}$ and $\zeta_{i}$ in the mixed basis,
respectively. For simplicity, we have defined the matrices $M_{RL}$ and $M_{LR}$ as combination of Left- and Right-components :
\begin{eqnarray}
M_{ij}^{RL}&:=&\frac{Y_{ij} \, v \, R + Z_{ij} \, u \, L}{\sqrt{2}}  \; ,
\nonumber \\
M_{ij}^{LR}&:=&\frac{Y_{ij} \, v \, L + Z_{ij} \, u \, R}{\sqrt{2}}  \; .
\end{eqnarray}

The Hermitian conjugate of (\ref{eqnuzeta}) is written below:
\begin{eqnarray}\label{eqnuzetaconj}
\bar{u}_{\nu_{i}}(p)\left( \, \slash{\!\!\!p} - m_{\nu_{i}} \, \right) -\bar{u}_{\zeta_{j}}(p) \, M_{ij}^{LR \, \dagger} &=& 0
\; ,
\nonumber \\
\bar{u}_{\zeta_{i}}(p)\left( \, \slash{\!\!\!p} - m_{\zeta_{i}} \, \right)- \bar{u}_{\nu_{j}}(p) \, M_{ij}^{RL \, \dagger} &=& 0
\; .
\end{eqnarray}
If we substitute $p^{\mu}$ by $p^{\prime\mu}$ in (\ref{eqnuzetaconj}), we can combine the Hermitian conjugate equations
with the equations (\ref{eqnuzeta}) to obtain the following tree-level Gordon-like decompositions:
\begin{eqnarray}\label{IDGordonnu}
&&
\bar{u}_{\nu_{i}}(p') \, \gamma^{\mu} \, u_{\nu_{i}}(p) = \bar{u}_{\nu_{i}}(p') \!\left( \frac{\ell^{\mu}}{2m_{\nu_{i}}}+i\, \frac{\sigma^{\mu\nu}q_{\nu}}{2m_{\nu_{i}}}\right)\!u_{\nu_{i}}(p)
\nonumber \\
&&
- \bar{u}_{\nu_{i}}(p') \! \left( \hspace{-0.25cm} \phantom{\frac{1}{2}}
\, \mu_{ij} \, \frac{\ell^{\mu}}{2m_{\nu_{i}}}+ \mu_{ij} \, i \, \frac{\sigma^{\mu\nu}q_{\nu}}{2m_{\nu_{i}}}
\right.
\nonumber \\
&&
\left.
+ \eta_{ij} \, \frac{q^{\mu}\gamma_{5}}{2m_{\nu_{i}}}
+ \eta_{ij} \, i \, \frac{\sigma^{\mu\nu}\ell_{\nu}}{2m_{\nu_{i}}} \, \gamma^{5} \hspace{-0.25cm} \phantom{\frac{1}{2}} \, \right) u_{\zeta_{j}}(p)
+\mbox{h. c.}
\, ,
\end{eqnarray}
and
\begin{eqnarray}\label{IDGordonzeta}
&&
\bar{u}_{\zeta_{i}}(p') \, \gamma^{\mu} \, u_{\zeta_{i}}(p) = \bar{u}_{\zeta_{i}}(p')\left( \frac{\ell^{\mu}}{2m_{\zeta_{i}}}+i\, \frac{\sigma^{\mu\nu}q_{\nu}}{2m_{\zeta_{i}}}\right) u_{\zeta_{i}}(p)
\nonumber \\
&&
- \bar{u}_{\zeta_{i}}(p') \!
\left( \hspace{-0.25cm} \phantom{\frac{1}{2}} \, \mu_{ij} \, \frac{\ell^{\mu}}{2m_{\zeta_{i}}}
+ \mu_{ij} \, i \, \frac{\sigma^{\mu\nu}q_{\nu}}{2m_{\zeta_{i}}}
\right.
\nonumber \\
&&
\left.
+\eta_{ij} \, \frac{q^{\mu}\gamma_{5}}{2m_{\zeta_{i}}}
+ \eta_{ij} \, i \, \frac{\sigma^{\mu\nu}\ell_{\nu}}{2m_{\zeta_{i}}} \, \gamma_{5} \phantom{\frac{1}{2}} \hspace{-0.25cm} \right) u_{\nu_{j}}(p)
+\mbox{h. c.} \; ,
\end{eqnarray}
where $q^{\mu}=p^{\mu}-p^{\prime \mu}$ is the photon's transfer momentum, and $\ell^{\mu}:=p^{\mu}+p^{\prime\mu}$
is the total $4$-momentum. These expressions yield the currents of neutrinos with the $\zeta_{i}$-fermions of the model, written in
momentum space, that we refer to as transition terms.
The coefficients $\mu_{ij}$ and $\eta_{ij}$ are matrix elements that depend on the Yukawa complex constant coupling and the
$u$-VEV scale :
\begin{eqnarray}
\mu_{ij} \!&=&\! \frac{1}{2\sqrt{2}} \, \frac{Z_{ij} \, u+Y_{ij} \, v}{m_{\zeta_{i}}+m_{\nu_{i}}} \; ,
\nonumber \\
\eta_{ij} \!&=&\! \frac{1}{2\sqrt{2}} \, \frac{Z_{ij} \, u-Y_{ij} \, v}{m_{\zeta_{i}}+m_{\nu_{i}}} \; .
\end{eqnarray}

Here, if we use that $m_{\zeta_{i}} \gg m_{\nu_{i}}$,
and $u \gg v$, so we can approximate $Y_{ij} \, v+Z_{ij} \, u \approx Z_{ij} \, u$, and the coefficients $\mu_{ij}$ and $\eta_{ij}$
are approximately equal, $\mu_{ij}\simeq\eta_{ij}$.
We also neglect terms of order $Z_{ij}^{2} \approx 0$ with respect to linear terms of $Z_{ij}$ in (\ref{IDGordonnu}) and (\ref{IDGordonzeta}).
We observe the emergence of TMDM for neutrinos and $\zeta_{i}$-fermions in both expressions (\ref{IDGordonnu}) and (\ref{IDGordonzeta}).
If we multiply these currents by $e \, A_{\mu}$, and using the representation $q_{\nu} \rightarrow i \, \partial_{\nu}$
for the photon momentum, the terms $\sigma^{\mu\nu}q_{\nu}$ have the following TMDM for the $\nu_{i}$-neutrino :
\begin{eqnarray}\label{muneutrino}
\mu_{ij}^{\, (\nu_{i})} \!&=&\! \frac{e \, \mu_{ij}}{2 \, m_{\nu_{i}}}=\frac{e \, Z_{ij} \, u}{4\sqrt{2} \, m_{\nu_{i}} \, m_{\zeta_{i}}}
\nonumber \\
&&
\hspace{-0.5cm}
\simeq \frac{Z_{ij}}{4\sqrt{2}} \, \left(\frac{2.8 \, \mbox{TeV}}{m_{\zeta_{i}}}\right) \,  \left( \frac{1 \, \mbox{MeV}}{m_{\nu_{i}}} \right) \, \mu_{B} \; .
\end{eqnarray}
We have a result at tree-level for Dirac neutrinos that depends on the $\nu_{i}$-neutrino mass and the $\zeta_{i}$-mass. This result allows
us to set up an  estimate for the $Z_{ij}$ coupling constant using the known result $|\mu^{(\nu)}| \, \lesssim \, 8 \, \times \, 10^{-15} \, \mu_{B}$ in the literature \cite{BellPRL2005}. We consider the mass spectrum of $m_{\nu_{i}} \, \sim \, 1 \, \mbox{eV}$ for neutrinos, the mass of $m_{\zeta_{1}}=0.5 \, \mbox{TeV}$ for the $\zeta_{1}$-fermion, and $u=2.8 \, \mbox{TeV}$. In so doing, the expression (\ref{muneutrino}) yields the $Z_{ij}$ coupling constant below :
\begin{eqnarray}
|Z_{ij}| \, \lesssim \, 7.5 \, \times \, 10^{-21} \; .
\end{eqnarray}
The TMDM for $\zeta_{i}$-hidden fermion is given by expression
\begin{eqnarray}
\mu_{ij}^{\, (\zeta_{i})}=\frac{e \, \mu_{ij}}{2 \, m_{\zeta_{i}}}
= \frac{Z_{ij}}{4\sqrt{2}} \, \left(\frac{2.8 \, \mbox{TeV}}{m_{\zeta_{i}}}\right) \,  \left( \frac{1 \, \mbox{MeV}}{m_{\zeta_{i}}} \right) \, \mu_{B}  \; .
\end{eqnarray}
Therefore, we can estimate the TMDM for the $\zeta_{1}$-hidden fermion :
\begin{eqnarray}
|\mu_{ij}^{\, (\zeta_{1})}| \, \lesssim \, 1.6 \, \times \, 10^{-26} \, \mu_{B} \; ,
\end{eqnarray}
where $\mu_{B}=3 \times 10^{-7} \, \mbox{eV}^{-1}$ is the Bohr magneton, in natural units $c=\hbar=1$.

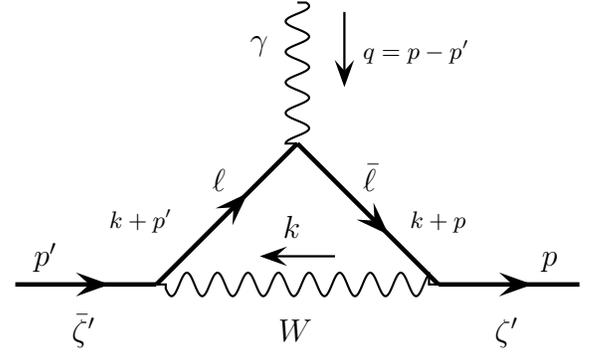
\begin{figure}[!h]
\begin{center}
\newpsobject{showgrid}{psgrid}{subgriddiv=1,griddots=10,gridlabels=6pt}
\begin{pspicture}(0,0.5)(8.8,5)
\psset{arrowsize=0.2 2}
\psset{unit=1.25}
%
%
%
%
\psline[linecolor=black,linewidth=0.6mm,ArrowInside=->,ArrowInsidePos=0.6](0.5,1)(2,1)
%
%
%
\psline[linecolor=black,linewidth=0.6mm,ArrowInside=->,ArrowInsidePos=0.6](5,1)(6.5,1)
%
%
%
%
%
%
%
\pscoil[coilaspect=0,coilarm=0.1,coilwidth=0.25,coilheight=1.3,linecolor=black](2,1)(5,1)
\psline[linecolor=black,linewidth=0.6mm,ArrowInside=->,ArrowInsidePos=0.6](2,1)(3.5,2.5)
\psline[linecolor=black,linewidth=0.6mm,ArrowInside=->,ArrowInsidePos=0.6](3.5,2.5)(5,1)
%
%
%
%
\pscoil[coilaspect=0,coilarm=0,coilwidth=0.25,coilheight=1.3,linecolor=black](3.5,2.5)(3.5,4)
%
%
%
%
%
%
%
%
\put(3.3,0.4){\large$W$}
\put(1.1,0.4){\large$\bar{\zeta}'$}
\put(5.6,0.4){\large$\zeta'$}
\put(2.6,2){\large$\ell$}
\put(4.2,2){\large$\bar{\ell}$}
\put(3,3.5){\large$\gamma$}
\put(4.2,3.4){$q=p-p^{\prime}$}
\psline[linecolor=black,linewidth=0.3mm]{->}(4,3.9)(4,3.1)
\put(0.7,1.2){\large$p^{\prime}$}
\put(4.7,1.6){$k+p$}
\put(1.5,1.6){$k+p^{\prime}$}
\put(6.1,1.2){\large$p$}
\psline[linecolor=black,linewidth=0.3mm]{<-}(3.1,1.3)(3.9,1.3)
\put(3.35,1.5){\large$k$}
\end{pspicture}
%
%
\caption{\scshape{The first one-loop contribution to the MDM of the $\zeta'_{i}$-fermion combining the
$W^{\pm}-\zeta'-\ell'$ vertex with the external photon.}}\label{Vertex1}
\end{center}
\end{figure}

The important contribution for the MDM
of $\zeta_{i}$ combines two external lines of $\zeta'_{i}$ with one external line of photon
in a one-loop diagram. This possible loop diagram emerges when we work in the mass basis $\zeta'_{i}$ due to the vertex
$W^{\pm}-\zeta'-\ell'$. This vertex depends on the $\theta$-mixing angle, and we believe that it must
correspond to a small effect. Obviously, the vertex does not show up whenever $\theta \, \rightarrow \, 0$.
The one-loop vertex is illustrated in figure (\ref{Vertex1}).
%
%
%
%
\begin{figure}[!h]
\begin{center}
\newpsobject{showgrid}{psgrid}{subgriddiv=1,griddots=10,gridlabels=6pt}
\begin{pspicture}(0,0.5)(8.8,5)
\psset{arrowsize=0.2 2}
\psset{unit=1.25}
%
%
%
%
\psline[linecolor=black,linewidth=0.6mm,ArrowInside=->,ArrowInsidePos=0.6](0.5,1)(2,1)
%
%
%
\psline[linecolor=black,linewidth=0.6mm,ArrowInside=->,ArrowInsidePos=0.6](5,1)(6.5,1)
%
%
%
%
%
%
%
%
\pscoil[coilaspect=0,coilarm=0.1,coilwidth=0.25,coilheight=1.3,linecolor=black](2,1)(3.5,2.5)
\pscoil[coilaspect=0,coilarm=0.1,coilwidth=0.25,coilheight=1.3,linecolor=black](3.5,2.5)(5,1)
%
%
%
%
%
%
\psline[linecolor=black,linewidth=0.6mm,ArrowInside=->,ArrowInsidePos=0.55](5,1)(2,1)
\pscoil[coilaspect=0,coilarm=0,coilwidth=0.25,coilheight=1.3,linecolor=black](3.5,2.5)(3.5,4)
%
%
%
%
%
%
%
%
\put(3.3,0.4){\large$\ell$}
\put(1.1,0.4){\large$\bar{\zeta}'$}
\put(5.4,0.4){\large$\zeta'$}
\put(2.25,2.1){\large$W^{-}$}
\put(4.25,2.1){\large$W^{+}$}
\put(2.75,3.5){\large$\gamma$}
\put(4.2,3.4){$q=p-p^{\prime}$}
\psline[linecolor=black,linewidth=0.3mm]{->}(4,3.9)(4,3.1)
\put(0.7,1.2){\large$p^{\prime}$}
\put(4.7,1.6){$k+p$}
\put(1.5,1.6){$k+p^{\prime}$}
\put(6.1,1.2){\large$p$}
%
%
\put(3.35,1.25){\large$k$}
%
%
%
%
\end{pspicture}
%
%
\caption{\scshape{The second one-loop contribution to the MDM of the$\zeta'_{i}$-fermion with the $W^{+}W^{-}$-photon vertex of the GSW model.}}\label{Vertex2}
\end{center}
\end{figure}
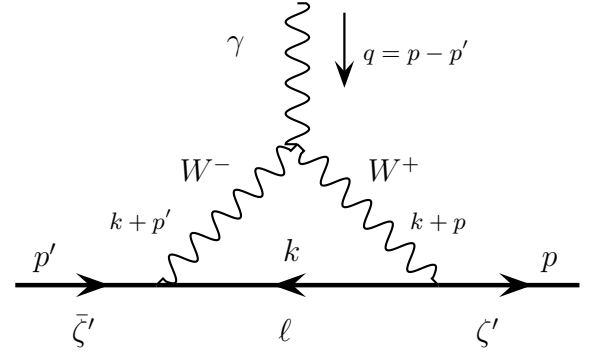

\noindent
Following the previous rules, the one-loop vertex (\ref{Vertex1}) can be calculated by the integral
\begin{eqnarray}\label{IntLambda}
\Lambda_{(1)ij}^{\mu}(p,p') \!\!&=&\!\! -\frac{g^2\theta^{2}}{8} \, V_{ik}V_{kj} \int \frac{d^4k}{(2\pi)^4} \, \gamma^{\alpha}\left(1- \gamma_{5}\right)
\nonumber \\
&&
\hspace{-1cm}
\times \,
\frac{i\left(\slash{\!\!\!k}+\slash{\!\!\!p}'+m\right)}{(k+p')^{2}-m^{2}}
\left(ie\gamma^{\mu}\right) \frac{i\left(\slash{\!\!\!k}+\slash{\!\!\!p}+m\right)}{(k+p)^{2}-m^{2}}
\nonumber \\
&&
\hspace{-1cm}
\times
\, \gamma_{\alpha}\left(1- \gamma_{5}\right) \frac{-i}{k^{2}-M_{W}^{2}} \; .
\end{eqnarray}
The second contribution comes from the combination of $W^{\pm}-\zeta'-\ell'$ interaction with the vertex $W^{\pm}$-photon
of the GSW model. It is illustrated in figure (\ref{Vertex2}). The Feynman rules yield the momentum-space loop integral below
\begin{eqnarray}\label{IntGamma}
\Lambda_{(2)ij}^{\mu}(p,p') \!\!&=&\!\! - \frac{g^2\theta^{2}}{8} \, V_{ik}V_{kj} \int \frac{d^4k}{(2\pi)^4} \, \gamma_{\alpha} \left(1-\gamma_{5}\right)
\nonumber \\
&&
\hspace{-1.5cm}
\times
\,
\frac{-i}{\left(k+p'\right)^{2}-M_{W}^{2}} V^{\mu\alpha\beta}\left(q,k+p',-k-p\right)
\nonumber \\
&&
\hspace{-1.5cm}
\times \,
\frac{-i}{\left(k+p\right)^{2}-M_{W}^{2}} \, \frac{i\left(\slash{\!\!\!k}+m\right)}{k^{2}-m^{2}} \, \gamma_{\beta}\left(1-\gamma_{5}\right) \; .
\end{eqnarray}
We have used the $W^{\pm}$-propagator in the Feynman gauge in the expressions (\ref{IntLambda}) and $(\ref{IntGamma})$, $V^{\mu\alpha\beta}$
sets the $W^{\pm}-\gamma$ vertex following the GSW model rules.
%
Well-known techniques to deal with Feynman integrals are introduced to calculate the finite part of these integrals and, then, the
contributions for the MDM of $\zeta_{i}$-fermions. The sum of these two contributions is denoted by $\Gamma^{\mu}=\Lambda_{(1)}^{\mu}+\Lambda_{(2)}^{\mu}$, so the finite part of $\Gamma^{\mu}$ at one-loop is written into the form
\begin{eqnarray}
J_{\zeta'_{i}(em)}^{\; \mu} \!\!&=&\!\!
\bar{u}_{\zeta'_{i}}(p') \left[ \phantom{\frac{1}{2}} \hspace{-0.25cm} f_{1}(q^2) \, \gamma^{\mu}+f_{2}(q^2) \gamma_{5} \, \gamma^{\mu}
\right.
\nonumber \\
&&
\hspace{-1cm}
\left.
+f_{3}(q^2) \, i \, \sigma^{\mu\nu}q_{\nu}+f_{4}(q^2) \, q^{\mu} \gamma_{5} \, \phantom{\frac{1}{2}} \hspace{-0.3cm}  \right]_{ij} \!\! u_{\zeta'_{j}}(p) \; ,
\end{eqnarray}
where $f_{1}$, $f_{2}$, $f_{3}$ and $f_{4}$ are the form factors of the previous diagrams to this order. The current conservation implies that
$q_{\mu}J_{\zeta'_{i}(em)}^{\mu}=0$; then, under this condition, we obtain the relation $f_{2}(q^2)=-q^2 \, f_{4}(q^2)/2M_{\zeta'_{i}}$.
Thereby, the EM-current of $\zeta'_{i}$ is reduced to expression
\begin{eqnarray}\label{currentEMzeta}
J_{\zeta'_{i}(em)}^{\; \mu}
\!\!&=&\!\! \bar{u}_{\zeta'_{i}}(p') \left[ \phantom{\frac{1}{2}} \hspace{-0.25cm} f_{1}(q^2) \, \gamma^{\mu}+f_{A}(q^2) \gamma_{5} \left(q^{\mu} \, \slash{\!\!\!q}-q^2 \, \gamma^{\mu} \right)
\right.
\nonumber \\
&&
\left.
+f_{3}(q^2) \, i \, \sigma^{\mu\nu}q_{\nu} \, \phantom{\frac{1}{2}} \hspace{-0.3cm} \right]_{ij} \! u_{\zeta'_{j}}(p) \; .
\end{eqnarray}
%

%
%
We have also used here the mass on-shell conditions for $\zeta'_{i}$-fermions : $\slash{\!\!\!p} \, u_{\zeta'_{i}}(p)=M_{\zeta'_{i}} \, u_{\zeta'_{i}}(p)$, $\bar{u}_{\zeta'_{i}}(p') \, \slash{\!\!\!p}'=\bar{u}_{\zeta'_{i}}(p') \, M_{\zeta'_{i}}$ and $p^{2}=p^{\prime2}=M_{\zeta'_{i}}^{2}$. The amplitudes $\bar{u}_{\zeta'_{i}}(p')$ and $u_{\zeta'_{i}}(p)$ stand for the plane wave solutions in the diagonal basis of $\zeta'_{i}$-fermions. The $f_{1}$-form factor is the contribution to electric charge given by
\begin{eqnarray}\label{fQ}
f_{1}(q^2)_{ij}=\frac{eg^{2}\theta^{2}}{32\pi^{2}} \, V_{ik}V_{kj} \int_{0}^{1} dx \, dy \, dz \, \delta(x+y+z-1)
\nonumber \\
\times \, \left[ \, \frac{M_{\zeta_{i}^{\prime}}^{2} \, z(4-z)-m^{2}}{M_{\zeta'_{i}}^{2} \, z(1-z)-M_{W}^{2}z-m^{2} \, z(1-z)+q^{2} xy}
\right.
\nonumber \\
\left.
-\frac{M_{\zeta_{i}^{\prime}}^{2} \, (1-z)(2+3z)}{M_{\zeta'_{i}}^{2} \, z(1-z)-M_{W}^{2}(1-z)-m^{2}z+q^{2} xy} \, \right] \, .
\hspace{0.5cm}
\end{eqnarray}
The second term in (\ref{currentEMzeta}) is known as the anapole (or toroidal momentum) term with the $f_{A}$-form factor that follows:
\begin{eqnarray}
f_{A}(q^2)_{ij}
= \frac{eg^{2}\theta^{2}}{32\pi^{2}} \, V_{ik}V_{kj} \int_{0}^{1} dx \, dy \, dz \, \delta(x+y+z-1)
\nonumber \\
\times \, \left[ \, \frac{2+2z-2(x-y)^2}{M_{\zeta'_{i}}^{2} \, z(1-z)-M_{W}^{2}z-m^{2} \, z(1-z)+q^{2} xy}
\right.
\nonumber \\
\left.
-\frac{3(1-z)-2(x-y)^{2}}{M_{\zeta'_{i}}^{2} \, z(1-z)-M_{W}^{2}(1-z)-m^{2}z+q^{2} xy} \, \right] \, .
\hspace{0.5cm}
\end{eqnarray}
The $f_{3}$-form factor is the contribution to the $\zeta'_{i}$-fermion MDM :
%
\begin{eqnarray}\label{fM}
f_{3}(q^2)_{ij} \!\!&=&\!\! \frac{e g^{2} \theta^{2}}{ 32 \pi^{2}} \, M_{\zeta'_{i}} \, V_{ik}V_{kj} \int_{0}^{1} dx \, dy \, dz \, \delta(x+y+z-1)
\nonumber \\
&&
\hspace{-1.3cm}
\times
\left[ \, \frac{z(1-z)}{M_{\zeta'_{i}}^{2} \, z(1-z)-M_{W}^{2}z-m^{2}z(1-z)+q^{2} xy}
\right.
\nonumber \\
&&
\hspace{-1.3cm}
\left.
-\frac{(1-z)(1/2-z)}{M_{\zeta'_{i}}^{2} \, z(1-z)-M_{W}^{2}(1-z)-m^{2}z+q^{2} xy} \, \right]  \, .
\hspace{0.2cm}
\end{eqnarray}
%
%
%
The on-shell condition for the external photon imposes $q^{2}=0$. Therefore, all the form factors depend on the masses of $\zeta'_{i}$-fermion,
$W^{\pm}$ and the lepton mass, where we take $M_{\zeta'_{i}} \, , \, M_{W} \gg m$ and also consider $M_{\zeta'_{i}} > M_{W}$ .
Under these conditions, the elements of the $f_{1}$-factor are
\begin{eqnarray}
f_{1}(0)_{ij}=\frac{e g^2 \theta^{2}}{16\pi^{2}} \, V_{ik}V_{kj} \, \left(1+\frac{M_{W}^2}{M_{\zeta'_{i}}^{2}} \right) \times
\nonumber \\
\times
\left[2-\left(1-\frac{M_{W}^{2}}{M_{\zeta'_{i}}^2}\right)\ln\left(\frac{M_{\zeta'_{i}}^2}{M_{W}^{2}}-1\right) \right] \; .
\end{eqnarray}
The interaction of $A^{\mu}$-field with
the $\zeta'_{i}$-current yields the form factor in the limit $q^{2} \, \rightarrow \, 0$
\begin{equation}
\bar{u}_{\zeta'_{i}}(p') \, \Gamma_{ij}^{\mu}(q) \, u_{\zeta'_{j}}(p) \, A_{\mu} \!\stackrel{q^{2} \rightarrow 0}{=}
-f_{3}(0)_{ij} \, \bar{u}_{\zeta'_{i}}(p') \, \vec{\sigma} \cdot \vec{{\bf B}} \, u_{\zeta'_{j}}(p) \; ,
\end{equation}
where the momentum transfer is represented in an operator form, $q_{\nu} \rightarrow i \, \partial_{\nu}$. We have considered the
EM tensor as $F^{\mu\nu}=\left(0, \epsilon^{ijk}B^{k} \right)$ with an external magnetic field, and $\sigma^{ij}=\varepsilon^{ijk}\sigma^{k}$.
We identify the elements of $\zeta'_{i}$-MDM as $\mu_{M \, ij}^{\, (\zeta'_{i})}=f_{3}(0)_{ij}$; thus, the $f_{3}(0)_{ij}$ form factor can be written
in terms of electron's mass and the Bohr magneton :
\begin{eqnarray}
\mu_{M \, ij}^{\, (\zeta'_{i})} \!&=&\!
\frac{3 G_{F} m_{e} M_{\zeta'_{i}}}{4\sqrt{2} \pi^{2}} \, \theta^{2} \, V_{ik}V_{kj} \, \frac{M_{W}^{2}}{M_{\zeta'_{i}}^{2}} \times
\nonumber \\
&&
\hspace{-0.5cm}
\times
\left[1-\frac{1}{3}\ln\left(\frac{M_{\zeta'_{i}}^{2}}{M_{W}^{2}}-1 \right)
+\frac{m_{W}^{2}}{M_{\zeta'_{i}}^{2}} \right] \, \mu_{B} \, .
\end{eqnarray}
This depends on the $M_{\zeta'_{i}}$-mass and on the ratio $M_{W}/M_{\zeta'_{i}}$, if we use $M_{W}=80 \, \mbox{GeV}$
and $M_{\zeta'_{1}}=0.5 \, \mbox{TeV}$, we have $M_{W}/M_{\zeta'_{1}}\simeq 0.14$, and $\theta\simeq -9 \, \times \, 10^{-8}$.
With these values, the MDM for the $\zeta'_{1}$-hidden fermion gets the one-loop contribution
\begin{eqnarray}
\mu_{M \, ij}^{\, (\zeta'_{1})}\simeq 1.2 \, \times \, 10^{-21} \; V_{ik} V_{kj} \, \mu_{B} \; .
\end{eqnarray}
%
%
%
%
%
Therefore, the $ij$-elements of the $\zeta'_{1}$-MDM depend on the $V$-matrix elements as given in (\ref{PMNSMatrix}). For example,
for the diagonal elements, $\mu_{M \, ii}^{\, (\zeta'_{1})}\simeq 1.2 \, \times \, 10^{-21} \; V_{ik} V_{ki} \, \mu_{B} $
with the implicit sum running over the $k$-index. For $i=1$, the element $\mu_{M \, 11}^{\, (\zeta'_{1})}$ depends on the cosines of mixing angles
$\theta_{12}$ and $\theta_{13}$, so we obtain the upper bound that follows:
\begin{eqnarray}
\mu_{M \, 11}^{\, (\zeta'_{1})}\simeq 1.2 \, \times \, 10^{-21} \, c_{12}^{\, 2} \, c_{13}^{\, 2} \; \mu_{B} \lesssim 1.2 \, \times \, 10^{-21} \, \mu_{B} \; .
\hspace{0.5cm}
\end{eqnarray}

\section{Concluding Comments}

We have here presented our efforts to discuss a model with an extra $U(1)_{Q'}$-factor that may describe a possible scenario for Particle Physics beyond the Standard Model (SM). The proposal is based on the gauge group $SU_{L}(2) \times U_{R}(1)_{Q}\times U(1)_{Q'}$, where the extra $U(1)_{Q'}$-factor introduces
a (new) massive neutral vector boson. The mass of the latter is generated upon the spontaneous symmetry breaking mechanism that defines an extra VEV
scale, beyond the electroweak VEV of $246 \, \mbox{GeV}$ associated with the usual Higgs field of SM. In our model, the Higgs sector displays two
scalar fields and the gauge symmetry allows interactions between them. This mechanism can be introduced in two ways :
in the first case, the SSB pattern takes place through a VEV scale-$u$, where $u > 246 \, \mbox{GeV}$, and, as consequence, we fix a mass
for the hypothetical $Z'$-boson around 2 TeV. Next, the SM Higgs acquires its VEV of $246 \, \mbox{GeV}$, breaks the
electroweak symmetry to give the $W^{\pm}$ and $Z$ masses. The sequence of SSBs is as follows:
\begin{eqnarray*}
SU_{L}(2) \! \times \! U_{R}(1)_{Q} \! \times \! U(1)_{Q'} \,
\stackrel{u \, \gg \, v}{\longmapsto} \, U_{em}(1) \; .
\end{eqnarray*}
The result for the mass of $Z'$ is
estimated by the ATLAS and CMS Collaborations as a possible particle at the $\mbox{TeV}$-scale.
The maximum VEV scale for $M_{Z'}=2 \, \mbox{TeV}$ is $u=2.8 \, \mbox{TeV}$, and it allows an estimation of the
mass of the new Higgs within the range $1.2 \, \mbox{TeV} < M_{F} < 3.7 \, \mbox{TeV}$.
Furthermore, the subgroup $U(1)_{K}$ also introduces a new family of fermions in the TEV-scale,
which we call $\zeta_{i}$-fermions $(i=1,2,3)$, that could be candidates to the DM particles.
It is a set of neutral heavy fermions associated with the VEV scale of $2.8 \, \mbox{TeV}$ and they
guarantee that model be free from the chiral anomaly. We use the masses for $\zeta_{i}$ in the range of
$0.5-1 \, \mbox{TeV}$, motivated by recent simulations of DM fermions scenario in the
CMS-Collaboration. Furthermore, these new fermions also mix with a Dirac's right-neutrino component.
This new mixing motivated us to investigate the MDM of the $\zeta_{i}$-fermions,
and the transition MDMs of the $\zeta_{i}$ mixed with neutrinos. Since $\zeta_{i}$
are neutral fermions, their MDMs depend on their masses and also on the mixing with the right-neutrinos.
At one-loop, the diagonal $\zeta_{1}$-MDM for mass of $0.5 \, \mbox{TeV}$ is estimated by the upper bound
$\mu_{M}^{(\zeta'_{1})} \lesssim 1.2 \, \times \, 10^{-21} \, \mu_{B}$.
%

An Electric Dipole Momentum (EDM) of the charged leptons must emerge at higher-loop orders.
This is an issue we are now investigating and we shall report on it elsewhere, in a further paper.
A unification scheme including $W'$- and $Z'$-bosons is in the framework of an Left-Right Symmetric Model
$SU_{L}(2)\times SU_{R}(2) \times U(1)_{Q} \times U(1)_{Q'}$ may also be the subject of a further investigation to be pursued
in the presence of fermionic DM.
%



%

%

\begin{thebibliography} {99}

\bibitem{Atlas} ATLAS Collaboration, {\it Phys. Rev.} D {\bf 90}, no. 5, 052005 (2014).

\bibitem{CMS} The CMS Collaboration, {\it JHEP} 1504, 025 (2015).

\bibitem{CMS20171} The CMS Collaboration, {\it Phys. Lett. B}, {\bf 2017}, {\it 773}, 563.

\bibitem{CMS20172} The CMS Collaboration, {\it J. High Energy Phys.}, {\bf 2017}, {\it 07}, 014.

\bibitem{CMS20173} The CMS Collaboration, {\it Phys. Rev.} D {\bf 97}, 092005 (2018).

\bibitem{DobrescuJHEP2015} Bogdan A. Dobrescu and Zhen Liu, {\it JHEP} (2015) 2015 : 118.

\bibitem{DobrescuPRL2015} Bogdan A. Dobrescu and Zhen Liu, {\it Phys. Rev. Lett.} {\bf 115} (2015) 211802.

\bibitem{DobrescuPRD2015} P. Coloma, B. A. Dobrescu and J. Lopez-Pavon, {\it Phys. Rev.} D {\bf 92} (2015) 115023.

\bibitem{DobrescuJHEP2016} Bogdan A. Dobrescu and Patrick J. Fox, {\it JHEP} (2016) 2016 : 047.

\bibitem{DevPRL2015} P. S. Bhupal Dev and R. N. Mohapatra, {\it Phys. Rev. Lett.} {\bf 115} (2015) 181803.

\bibitem{Patra2016} S. Patra, W. Rodejohann and C. E. Yaguna {\it JHEP} (2016) 2016 : 076.

\bibitem{HongGuPRD2017} Pei-Hong Gu and Rabindra N. Mohapatra, {\it Phys. Rev.} D {\bf 96}, 055011 (2017).

\bibitem{Dev2017} P. S. Bhupal Dev, Rabindra N. Mohapatra and Yongchao Zhang,
{\it Journal of Physics : Conf. Series} {\bf 873} (2017) 012029.

\bibitem{LangackerRMP2009} Paul Langacker, {\it Reviews of Modern Physics}, {\bf 81} (2009) 1199-1228.

\bibitem{Kanemura2011} Shinya Kanemura, Osamu Seto and Takashi Shimomura, {\it Phys. Rev.} D {\bf 84} (2011) 016004.

\bibitem{MorettiarXiv2017} E. Accomano {\it et al}, {\it JHEP} (2016) 2016 : 86.


\bibitem{MJNeves2017Annalen} M. J. Neves and J. A. Helay\"el-Neto, {\it Annalen Der Physik} {\bf 2017}, 1700112.

\bibitem{BellPRL2005} Nicole F. Bell {\it et. al.},
{\it Phys. Rev. Lett.} {\bf 95} (2005) 151802.

\bibitem{BellPLB2006} Nicole F. Bell {\it et. al.}, {\it Phys. Lett.} B 642 (2006) 377-383.

\bibitem{Bhattacharya2004} Kaushik Bhattacharya and Palash B Pal, {\it Proc Indian Natn Sci Acad}, {\bf 70}, A No. 1, January 2004, pp. 145–161.

\bibitem{Balantekin2006} A. B. Balantekin, AIP Conference Proceedings {\bf 847}, 128 (2006).

\bibitem{StudenikinJP2016} Alexander Studenikin, {\it Journal of Physics} : Conference Series {\bf 718} (2016) 062076.

\bibitem{PDG2016} C. Patrignani {\it et. al.} (Particle Data Group), {\it Chin. Phys.} C 40 100001 (2016).

\bibitem{ArakiPRL2005} KamLAND Collaboration, {\it Phys. Rev. Lett.}, {\bf 2005}, {\it 94}, 081801.

\end{thebibliography}
\end{document}